\documentclass[showpacs,preprintnumbers,amsmath,amssymb]{revtex4}
\usepackage{amsmath}
\usepackage{graphicx}
\usepackage{subfig}
\usepackage{hyperref}
\usepackage{amssymb}
\usepackage[english]{babel}
\usepackage{epsfig}
\usepackage{wasysym}

\begin{document}
\title{ Regular Black Holes in Rainbow Gravity\\}
\author{Ednaldo L. B. Junior$^{(a)}$}\email{ednaldobarrosjr@gmail.com}
\author{Manuel E. Rodrigues$^{(b, c)}$}\email{esialg@gmail.com}
\author{Marcos V. de S. Silva$^{(c)}$}\email{marco2s303@gmail.com}

\affiliation{$^{(a)}$ Faculdade de Engenharia da Computa\c{c}\~{a}o, 
Universidade Federal do Par\'{a}, Campus Universit\'{a}rio de Tucuru\'{\i}, CEP: 
68464-000, Tucuru\'{\i}, Par\'{a}, Brazil}
\affiliation{$^{(b)}$ Faculdade de Ci\^{e}ncias Exatas e Tecnologia, 
Universidade Federal do Par\'{a}\\
Campus Universit\'{a}rio de Abaetetuba, CEP 68440-000, Abaetetuba, Par\'{a}, 
Brazil} 
\affiliation{$^{(c)}$ Faculdade de F\'{\i}sica, PPGF, Universidade Federal do 
 Par\'{a}, 66075-110, Bel\'{e}m, Par\'{a}, Brazil}
\begin{abstract}
In this work, we consider that in energy scales greater than the Planck energy, the geometry, fundamental physical constants, as charge, mass, speed of light and Newtonian constant of gravitation, and matter fields will depend on the scale. This type of theory is known as Rainbow Gravity. We coupled the nonlinear electrodynamics to the Rainbow Gravity, defining a new mass function $M(r,\epsilon)$, such that we may formulate new classes of spherically symmetric regular black hole solutions, where the curvature invariants are well-behaved in all spacetime. The main differences between the General Relativity and our results in the the Rainbow gravity are: a) The intensity of the electric field is inversely proportional to the energy scale. The higher the energy scale, the lower the electric field intensity; b) the region where the strong energy condition (SEC) is violated decrease as the energy scale increase. The higher the energy scale, closer to the radial coordinate origin SEC is violated.  

\end{abstract}

\maketitle


\section{Introduction}
\label{sec1}

One of the most interesting predictions of Einstein theory are black hole solutions. These objects have been challenging physicists since Karl Schwarzschild, in 1915, first predicted their existence by solving the Einstein equations for the vacuum. The Schwarzschild solution, although elegant, had the characteristic that brought out the name ``black hole'', which is the existence of a region of spacetime where it is only possible to enter, not to scape from, delimited by the so-called event horizon. This region has a singularity in the origin of the radial coordinate. Since that, other solutions arose in the literature, such as the solution from the first coupling with matter, the Reissner \cite{35} and Nordstrom solution \cite{36}, where the coupling with matter was done through Maxwell electromagnetism, resulting in a spherically symmetric, static and charged solution, known as the Reissner-Nordstrom black hole. Another well-known solution of Einstein equations, which can be understood as a generalization of the Schwarzschild metric, is the Kerr solution \cite{kerr}, where rotation effects are considered in a spacetime without matter sources. Since then, numerous works have been done in search of new solutions to Einstein equations. However, these structures remained hidden from the eyes of the scientific community, until April 2019, the Event Horizon Telescope collaboration, led by John Wardle, revealed to the world the first image of a black hole \cite{Horizon}, further contributing to the interest in these objects. 

One of the great challenges of theoretical physics is the search for a unification of gravitation with quantum theory. Many attempts at unification have been, and are being proposed \cite{unifi}, however, none have the consistency necessary for a complete description of the theory. One of the main motivations for unifying gravity with quantum theory is the violation of Lorentz invariance, which we can consider as an essential requirement for formulating a quantum theory of gravity. Other feature of unification theories is that they predict a maximum energy value on the order of the Planck scale at which a particle can reach, in other words, if a particle is described by the standard model then it must obey the upper limit of Planck energy, $5.10^{19} eV$, value experimentally confirmed by J. Abraham et al \cite{Abraham}. An alternative to Lorentz symmetry violation is the modification of the standard energy-momentum dispersion ratio at the ultraviolet limit, this has already been observed in some theories like Horava-Lifshitz \cite{Horava}, loop quantum gravity \cite{loop}, discreteness spacetime \cite{disc} and  doubly special relativity \cite{Espdupla}. To the doubly special relativity, the dispersion relation may be written as $E^2f(\epsilon)^2-p^2g(\epsilon)^2=m^2$ where $f(\epsilon)$ and $g(\epsilon)$ are functions of  $\epsilon=E/E_P$ with $E$ being the energy of the particle used to analyze the spacetime and $E_P$ is the Planck energy. The functions $f(\epsilon)$ and $g(\epsilon)$ are known as rainbow functions and have phenomenological motivations \cite{Mfg}. The standard energy dispersion relations is recurved in the limit $\lim_{\epsilon\rightarrow 0}f(\epsilon)=\lim_{\epsilon\rightarrow 0}g(\epsilon)=1$, known as infrared limit. The linear Lorentz transformations break the invariance of the theory, however, nonlinear Lorentz transformations are those that hold the theory invariant \cite{remo2}, in addition, not only the speed of light is a constant, but also Planck's energy, and it is impossible for a particle to reach energies greater than this limit.

Magueijo and Smolin \cite{Smolin} proposed a generalization from the doubly special relativity to what they call Rainbow Gravity, where the spacetime is represented by a family of parameters in the metric that is parameterized by $\epsilon$, causing the spacetime geometry to depend on the energy of the particle that is being used for test it, thus creating a rainbow of metric. Since then some researchers have considered the Magueijo and Smolin theory in the study of black holes \cite{Zhang}-\cite{Eslam} and in generalized theories of gravity \cite{remo1}-\cite{Momenia3}. In the context of nonlinear electrodynamics (NED), we can highlight the M. Momennia et al. work \cite{Momennia}, which shows that there is an essential singularity covered by an event horizon and verified the validity of the first law of thermodynamics in the presence of rainbow functions. In \cite{Momennia2}, the authors obtained exact black hole solutions in the Born-Infeld-dilaton gravity with a energy dependent Liouville-type potential, as well as the thermodynamic quantities. In \cite{Faizal}, it has been shown that the inclusion of Gauss-Bonnet gravity in the context of Rainbow Gravity can significantly alter the constraints needed to obtain non-singular universes. One application of Rainbow Gravity is the black hole thermodynamics. Due to the energy scale dependence on the spacetime metric, entropy and temperature also have the same dependence \cite{Panah}. Thus, the thermodynamics of these new solutions are altered showing the influence of the energy scale on the thermal system. In the NED context, Dehghani showed the thermal fluctuations of AdS black holes in three-dimensional rainbow gravity \cite{MDeh}.

A large class of regular black hole solutions found in the literature follows the idea that singularity is replaced by a regular distribution of matter \cite{39}. In usual General Relativity, NED produces black holes where singularity is eliminated by a regular field distribution that covers the central core of the black hole. We must consider an important feature of the regular solutions that is linked to so-called energy conditions, all regular black hole solution with spherical symmetry violate, in some region, the strong energy condition (SEC) \cite{SEC}, but not necessarily the dominant (DEC) and weak (WEC) energy conditions, as we have in the Bardeen solution \cite{barden}.

Our focus in this work is to get black hole solutions that are regular in origin and investigate how this can be changed by the $ \epsilon $ power scale parameter. In addition, we are interested in how these new solutions behave in the energy conditions.

\section{Regular Black Holes}
\label{sec2}
In order to study the Rainbow gravity, General Relativity must be reformulated considering a dependency on the energy scale and now should be describe by a set of parameters in the connection, $\Gamma^{\mu}_{\;\;\alpha\beta}(\epsilon)$, Riemann tensor, $R(\epsilon)^\sigma_{\,\,\,\mu\nu\lambda}$, which are construct using a rainbow metrics, and in the stress-energy tensor $T(\epsilon)_{\mu\nu}$. So that, the modified Einstein equations to the Rainbow gravity are given by
\begin{eqnarray}
G(\epsilon)_{\mu\nu}=\kappa(\epsilon)^2 T(\epsilon)_{\mu\nu},
\end{eqnarray}
where $\kappa(\epsilon)^2=8\pi G(\epsilon)/c^4(\epsilon)$ is a coupled constant and $G(\epsilon)=h^2(\epsilon)G_N$, with $G_N$ being the Newtonian gravitational constant, where $h(\epsilon)=1/[1+(G_N/137)\epsilon^2]$ and $c(\epsilon)=g(\epsilon)/f(\epsilon)$ \cite{Smolin,Anber}. The general line element, that describes spherically symmetric configurations and depends on $\epsilon$, is written as 
\begin{eqnarray}
ds^2=\frac{e^{a(r)}}{f(\epsilon)^2}dt^2-\frac{e^{b(r)}}{g(\epsilon)^2}dr^2-\frac{r^2}{g(\epsilon)^2}(d\theta^2+\sin^2\theta d\phi^2)\label{M1},
\end{eqnarray}
where $e^{a(r)}$ and $e^{b(r)}$ are general functions of the radial coordinate and the dispersion relations proposed in \cite{fg}, $f(\epsilon)$ and $g(\epsilon)$, 
are $f(\epsilon)=g(\epsilon)=1/(1+\lambda\epsilon)$, where $\lambda$ is a constant parameter (in this case $c(\epsilon)$ is a constant). The curvature scalar is given by
\begin{eqnarray}
R(\epsilon)=e^{-b}g(\epsilon)^2 \left[a''+\left(\frac{a'}{2}+\frac{2}{r}\right)(a'-b')+\frac{2}{r^2}\right]-2\frac{g(\epsilon)^2}{r^2}\,.\label{esc1}
\end{eqnarray}

The coupled with NED is made through the stress-energy tensor $T(\epsilon)^{\,\,\,\nu}_\mu$, that is 
\begin{eqnarray}
T(\epsilon)^{\,\,\,\mu}_\nu= L(F)\delta^\mu_\nu-\frac{\partial L(F)}{\partial F}F_{\nu\sigma}F^{\mu\sigma}\,.\label{TEM} 
\end{eqnarray}
We define the the 4-potential as being a 1-form $A=A_{\mu}dx^{\mu}$. The Maxwell tensor is a 2-Form defined as $\hat{F}=(1/2)F_{\mu\nu}dx^{\mu}\wedge dx^{\nu}$, where the components are $F_{\mu\nu}=\nabla_{\mu}A_{\nu}-\nabla_{\nu}A_{\mu}$. Them, the electromagnetic scalar becomes $F=(1/4)F^{\mu\nu}F_{\mu\nu}$. If we consider that the source is only electrically charged and has spherical symmetry, the only non-zero and independent component os the Maxwell tensor is $F_{10}(r)$ \cite{plebanski}. The modified Maxwell equations are
\begin{equation}
\nabla _{\mu}\left[F^{\mu\nu}\frac{\partial L}{\partial F}\right]=\partial_{\mu}\left[\sqrt{-g}F^{\mu\nu}\frac{\partial L}{\partial F}\right]=0\label{mMaxeq}\,,
\end{equation}
whose the solution, to the line element \eqref{M1}, is given by \footnote{In \cite{balart}, the electric field differ from the result obtained here, in the context of General Relativity, since the authors consider that the solution to the modified Maxwell equations is written as $F^{10}\partial L/\partial F=q/(4\pi r^2)$.}
\begin{eqnarray}
F^{10}(r,\epsilon)=f(\epsilon) g^3(\epsilon)\frac{q(\epsilon)}{r^2}e^{-(a+b)/2}\left(\frac{\partial L}{\partial F}\right)^{-1}\,,\label{F10}
\end{eqnarray}
where $q(\epsilon)$ is a function associated to the electric charge \cite{carga} as $q(\epsilon)=Q\sqrt{k(\epsilon)}$, with $Q$ being the electric charge and $k(\epsilon)=1/(1-\frac{Q^2\log\epsilon}{6\pi^2})$ \cite{livro}.

Using \eqref{M1}, \eqref{F10} and \eqref{TEM}, with $L_F=\partial L/\partial F$, the components of the field equations are
\begin{eqnarray}
\frac{g(\epsilon)^2}{r^2}\left[e^{-b}(rb'-1)+1\right]=\kappa(\epsilon)^2\left(L+g(\epsilon)^4\frac{q(\epsilon)^2}{r^4}L^{-1}_{F}\right)\,,\label{eq1}\\
-\frac{g(\epsilon)^2}{r^2}\left[e^{-b}(ra'+1)-1\right]=\kappa(\epsilon)^2\left(L+g(\epsilon)^4\frac{q(\epsilon)^2}{r^4}L^{-1}_{F}\right)\,,\label{eq2}\\
-\frac{g(\epsilon)^2e^{b}}{4r}\left[\left(2+a'r\right)\left(a'-b'\right)+2ra''\right]=\kappa(\epsilon)^2L\,.\label{eq3}
\end{eqnarray}
Through the condition $T^{\,\,\,0}_{0}=T^{\,\,\,1}_{1}$, we may choose $a(r)=-b(r)$. Identifying $T_{\,\,\,0}^{0}=\rho$, $T_{\,\,\,1}^{1}=-p_r$ and $T_{\,\,\,2}^{2}=T_{\,\,\,3}^{3}=-p_t$ where $\rho$ is the energy density and $p_r$ and $p_t$ are, respectively, the radial and tangential pressures. To the line element \eqref{M1}, we get 
\begin{eqnarray}
&&\rho(r, \epsilon)=\frac{e^{-b}g(\epsilon)^6}{r^28\pi G_N f(\epsilon)^4h(\epsilon)^2}\left(rb'+e^{b}-1\right)\,,\\ \label{rho}
&&p_r(r,\epsilon)=-\rho(r,\epsilon)\,,\label{pr1}\\
&&p_t(r,\epsilon)=-\frac{e^{-b}g(\epsilon)^6}{16r\pi G_Nf(\epsilon)^4h(\epsilon)^2}\left[rb'^2-rb''-2b'\right]\,.\label{pr2}
\end{eqnarray}
The energy conditions will be given by \cite{econdi}:
\begin{eqnarray}
&&SEC=\rho+p_r+2p_t\geqslant 0\,,\label{SEC}\\
&&WEC_{1,2}=\rho+p_{r,t}\geqslant 0\,,\label{WEC}\\
&&WEC_3=\rho\geqslant 0\,,\label{WE3}\\
&&DEC_{2,3}=\rho-p_{r,t}\geqslant 0\,,\label{DEC}
\end{eqnarray}
where $DEC_1\equiv WEC_3$ and $NEC_{1,2}\equiv WEC_{1,2}$. 

Using the line element $\eqref{M1}$, we may choose different models of mass functions to construct regular black holes solutions. To do that, we introduce the mass function through the coefficient of the metric $g_{00}$ as
\begin{eqnarray}
e^{a(r)}=1-2\frac{G(\epsilon)}{c^2(\epsilon)}\frac{M(r, \epsilon)g(\epsilon)}{r}\,.\label{ea}
\end{eqnarray}
If the solution is regular in $r=0$, $M(r, \epsilon)$ must satisfy the condition $\lim_{r\rightarrow 0}\left[M(r,\epsilon)/r\right]\rightarrow 0 $. 

From the equations \eqref{eq1}-\eqref{eq3}, $L$ and $L_F$, in terms of $M(r,\epsilon)$, are
\begin{eqnarray}
L=G_N f^2(\epsilon) g(\epsilon) h^2(\epsilon)\frac{M''(r,\epsilon)}{r\kappa^2(\epsilon)}\,,\label{LM1}\\
L_F=\frac{q(\epsilon)^2g^3(\epsilon)\kappa^2(\epsilon)}{r^2\,G_Nf^2(\epsilon)h^2(\epsilon)\left(-2M'(r,\epsilon)+M''(r,\epsilon)\right)}\,.\label{LM2}
\end{eqnarray}
The Lagrangian density $L$ and its derivate $L_F$ must satisfy the relation 
\begin{eqnarray}
L_F=\frac{\partial L}{\partial r}\left(\frac{\partial F}{\partial r}\right)^{-1}\,.\label{Lf}
\end{eqnarray}
In order to do that, we need the electric field, that is
\begin{eqnarray}
F^{10}(r,\epsilon)=G_Nf^3(\epsilon) h^2(\epsilon)\frac{\left(2M'(r,\epsilon)-rM''(r,\epsilon)\right)}{q(\epsilon)\,g(\epsilon) \kappa^2(\epsilon)}\,.\label{F102}
\end{eqnarray}
Using the relation $F=-e^{a+b}[F^{10}(r,\epsilon)]^2(2f^2(\epsilon)g^2(\epsilon))^{-1}$, it's possible to show that \eqref{Lf} is satisfied. 

We will consider some model to the mass function in order to obtain regular solutions that depend on the $\epsilon$.

\subsection{Culetu-type solution}

The solution proposed by Culetu in \cite{culetu} is described by the line element 
\begin{eqnarray}
ds^2=\left(1-\frac{2m}{r}e^{-p_0/r}\right)dt^2-\frac{1}{1-\frac{2m}{r}e^{-p_0/r}}dr^2-r^2\left(d\theta^2+\sin^2\theta  d\phi^2\right),
\end{eqnarray} 
where $p_0$ is a parameter associated with the charge $q$ and mass $m$, such that $p_0=q^2/2m$ generates regular solution in GR, which is asymptotically Reissner-Nordström. Let's extend the Culetu solution to the Rainbow Gravity using the line element \eqref{M1} with \eqref{ea} and 
\begin{eqnarray}
M(r,\epsilon)=me^{-Q^2 k(\epsilon)g(\epsilon)/(2mr)}\label{Model1},
\end{eqnarray}
with $m$ being the black hole mass. This model was consider and analyzed in \cite{balart} to the case $g(\epsilon)=k(\epsilon)=1$. 

\begin{figure}[h]
\centering
\begin{tabular}{rl}
\includegraphics[height=5cm,width=8cm]{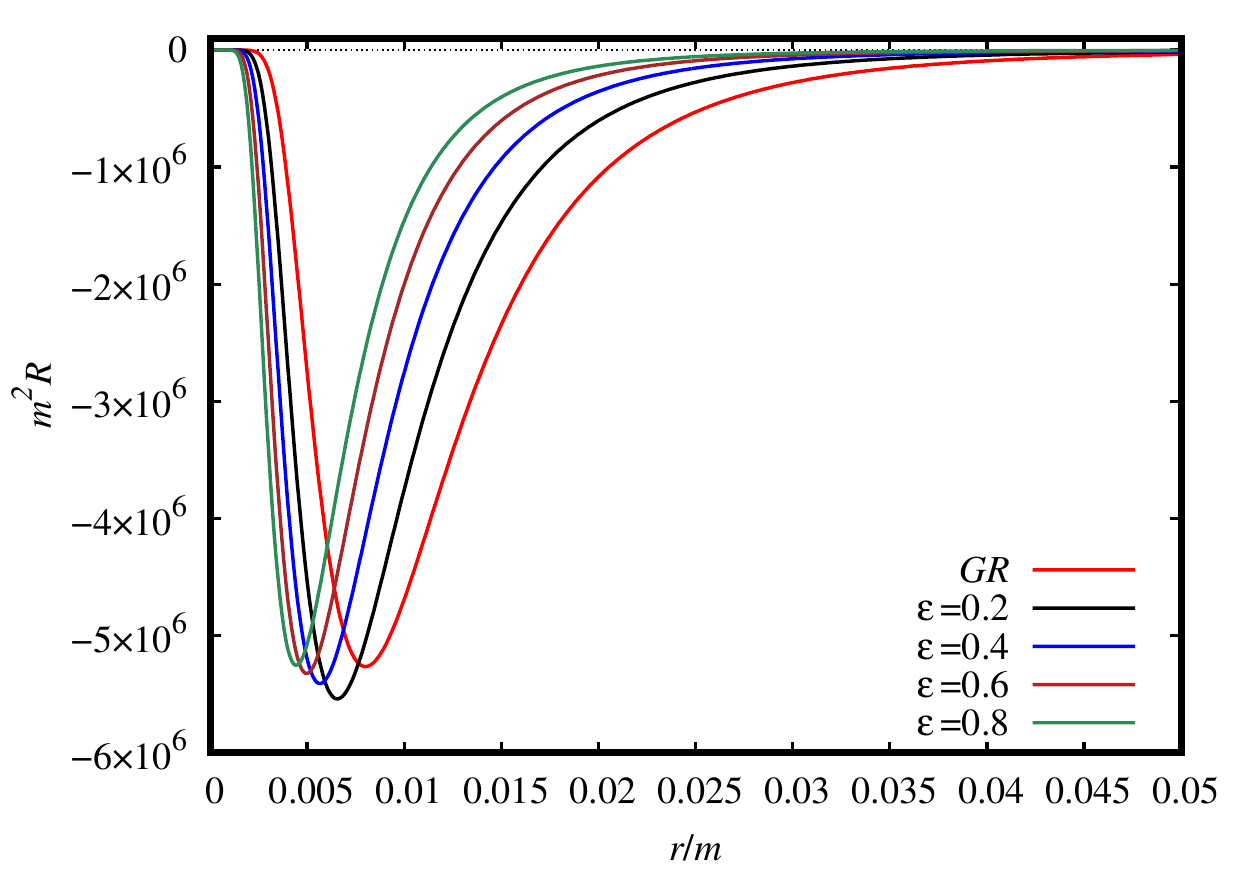}&\includegraphics[height=5cm,width=8cm]{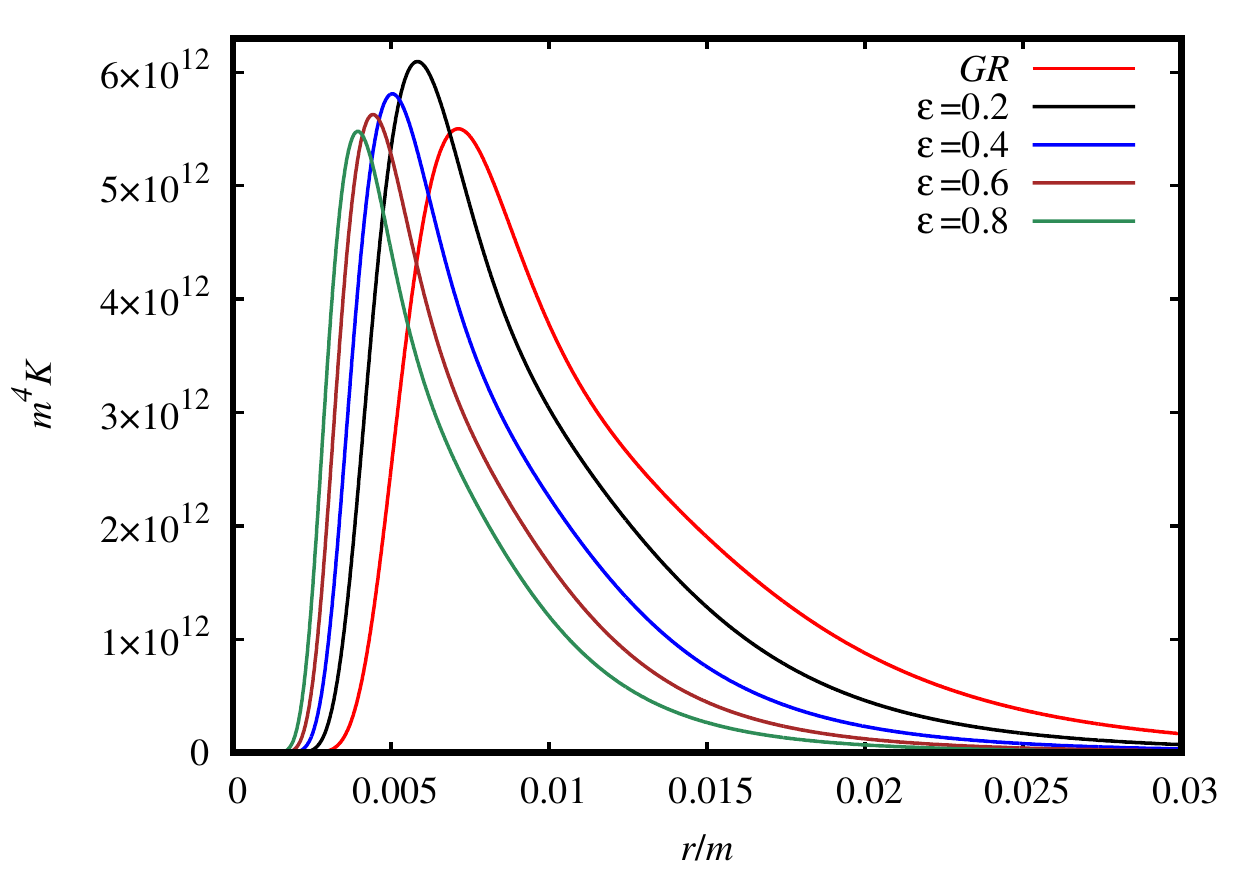}\\
\end{tabular}
\caption{\scriptsize{Graphical representation of the functions $R(r, \epsilon)$ and $K(r,\epsilon)$, to $m=10Q$, $Q=0.8$, $\lambda=1$, and $G=1$ to different values of energy with the result from GR.} }
\label{fig2}
\end{figure}

Using \eqref{Model1} and \eqref{ea} in \eqref{M1}, we may get the curvature scalar \eqref{esc1} to this model, that is 
\begin{eqnarray}
R(r,\epsilon)=-\frac{e^{-Q^2\frac{g(\epsilon)k(\epsilon)}{2mr}}G_N\,Q^4\,f^2(\epsilon)g^3(\epsilon)h^2(\epsilon)k^4(\epsilon)}{2m r^5} \,.\label{R1}
\end{eqnarray}
We also calculate the Kretschmann scalar $K=R^{\alpha\beta\mu\nu}R_{\alpha\beta\mu\nu}$, which results in 
\begin{eqnarray}
K(r,\epsilon)&=&\frac{e^{-\frac{Q^2g(\epsilon)k(\epsilon)}{mr}}}{4m^2r^{10}}G^2_N\,f(\epsilon)^4g(\epsilon)^2h(\epsilon)^4\nonumber\\
&&\times\left[192m^4r^4+Q^2g(\epsilon)k(\epsilon)(Q^2g(\epsilon)k(\epsilon)-4mr)\left(48m^2r^2+Q^2g(\epsilon)k(\epsilon)(Q^2g(\epsilon)k(\epsilon)-12mr)\right)\right]\,.\label{kre}
\end{eqnarray}
The scalars \eqref{R1} and \eqref{kre} are regular in all spacetime to $m>0$, $f(\epsilon)>0$, $g(\epsilon)>0$, $h(\epsilon)>0$ and $k(\epsilon)>0$, once in the limits $r\rightarrow 0$ and $r\rightarrow \infty$, both $R(r,\epsilon)$ and $K(r,\epsilon)$ do not present divergences. In Fig. \ref{fig2}, the behavior of the scalars are exhibit and we compare to different values of $\epsilon$.
The scalar $R(r,\epsilon)$ and $K(r,\epsilon)$, that depend on the energy $\epsilon$, are regular in all spacetime and zero in the infinity, which characterize a regular and asymptotically flat solution.

\begin{figure}[h]
\centering
\begin{tabular}{rl}
\includegraphics[height=5cm,width=8cm]{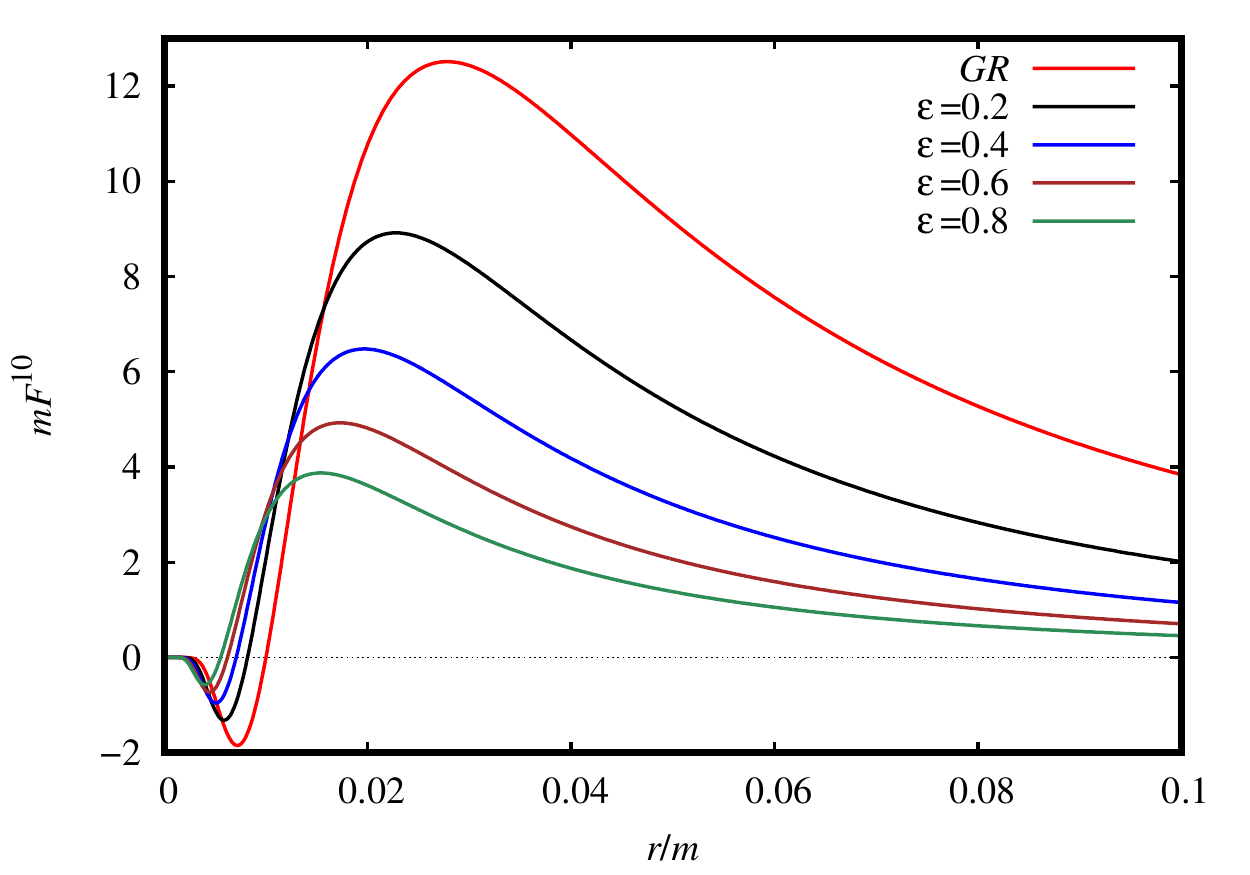}
\end{tabular}
\caption{\scriptsize{Intensity of the electric field to $m=10Q$, $Q=0.8$, $\lambda=1$, $G=1$ comparing different values of energy with the GR case.} }
\label{fig1}
\end{figure}

\begin{figure}[h]
\centering
\begin{tabular}{rl}
\includegraphics[height=5cm,width=8cm]{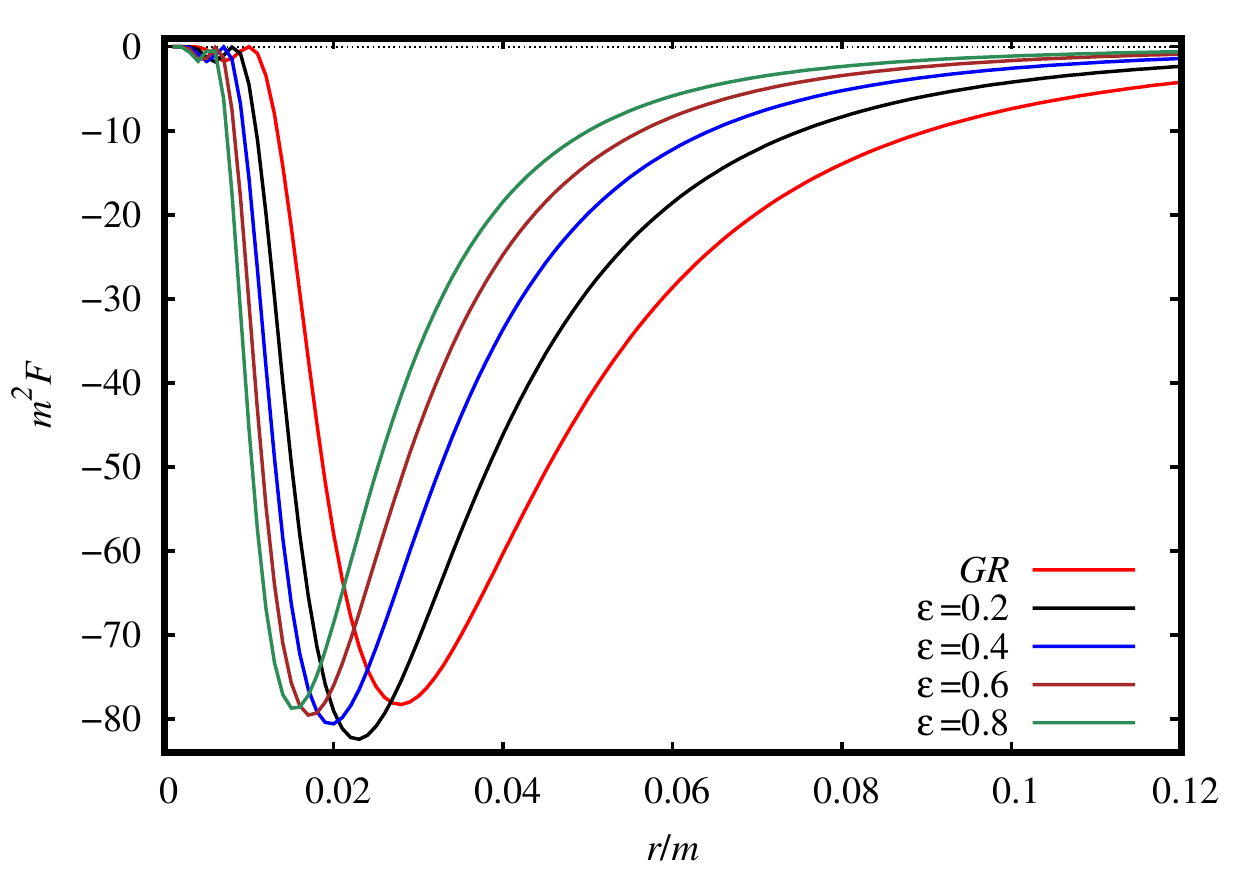}&\includegraphics[height=5cm,width=8cm]{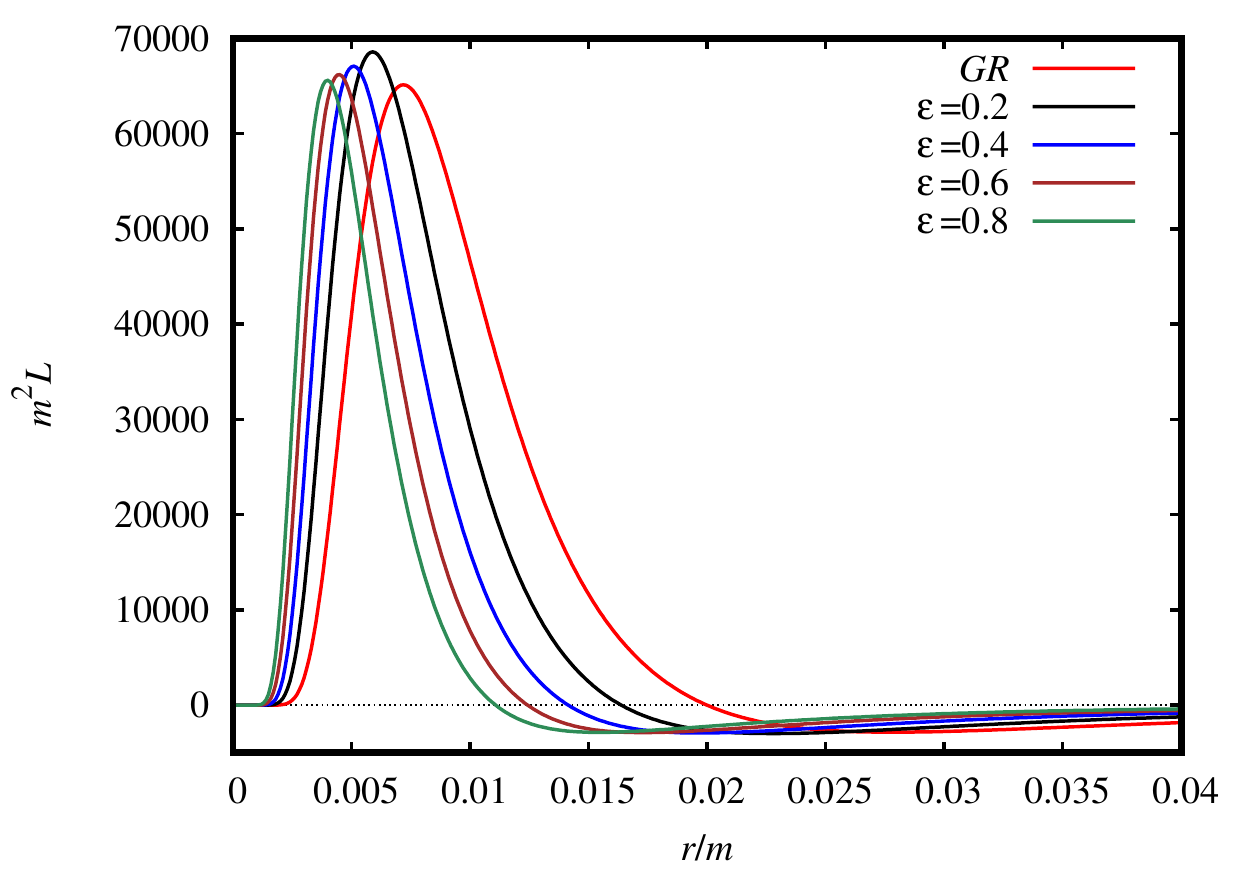}\\
\includegraphics[height=5cm,width=8cm]{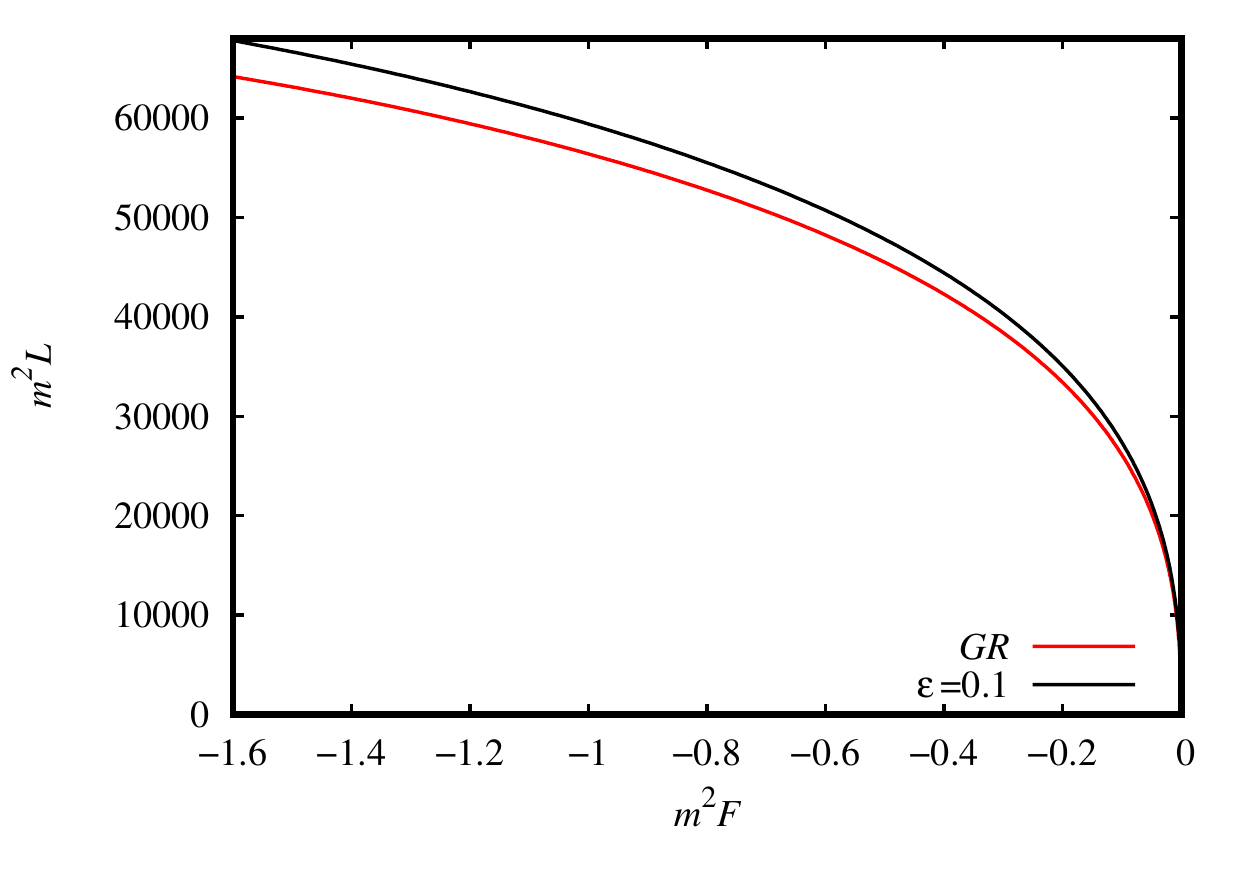}
\end{tabular}
\caption{\scriptsize{Graphical representation of the functions $F(r, \epsilon)$ (left) and $L(r,\epsilon)$ (right) with $m=10Q$, $Q=0.8$, $\lambda=1$, $G=1$ to different $\epsilon$ and GR limit. In the bottom graph, we show the difference between $L(F)$ to GR and Rainbow gravity with $\epsilon=0.1$.} }
\label{fig3}
\end{figure}

Substituting \eqref{Model1} in \eqref{F102}, the electric field becomes 
\begin{eqnarray}
F^{10}(r,\epsilon)=\frac{e^{-\frac{Q^2g(\epsilon)k(\epsilon)}{2mr}}G_N\,Q^2\,(8mr-Q^2g(\epsilon)k(\epsilon))}{4mr^3 q(\epsilon) \kappa(\epsilon)^2}f^3(\epsilon)g^2(\epsilon)\,.\label{E10fg}
\end{eqnarray}  
The electric field behavior is exhibit in Fig. \ref{fig1} to different values of $\epsilon$ and we also compare to the GR result.
The intensity of the electric field is always regular and presents a minimum and a maximum. The intensity of \eqref{E10fg} decreases as $\epsilon$ increases. Since we have $F^{10}$, we may evaluate $F(r,\epsilon)$ and the Lagrangian density $L(r,\epsilon)$. Using \eqref{E10fg}, $F$ is given by 
\begin{eqnarray}
F(r,\epsilon)=-\frac{e^{-\frac{Q^2g(\epsilon)k(\epsilon)}{mr}}G_N^2\,Q^2(Q^2g(\epsilon)k(\epsilon)-8mr)^2h(\epsilon)^4f(\epsilon)^4k(\epsilon)^2}{32m^2r^6q(\epsilon)^2\kappa(\epsilon)^4}\,,\label{F}
\end{eqnarray} 
and from \eqref{LM1} and \eqref{Model1}, the Lagrangian density to the nonlinear electrodynamics is
\begin{eqnarray}
L(r,\epsilon)=-\frac{e^{-\frac{Q^2g(\epsilon)k(\epsilon)}{2mr}}G_N\,Q^2(Q^2g(\epsilon)k(\epsilon)-4mr)f(\epsilon)^2h(\epsilon)^2k(\epsilon)}{4mr^5\kappa(\epsilon)^2}\,.\label{L}
\end{eqnarray}
The behavior of \eqref{F} (right) and \eqref{L} (left), in terms os the radial coordinate, as well $L(F)$ (bottom) is analyzed in Fig. \ref{fig3} and compared to the GR result.
It's clear the nonlinearity of the electromagnetic theory and the regularity of these functions.

\begin{figure}[h]
\centering
\begin{tabular}{rl}
\includegraphics[height=5cm,width=8cm]{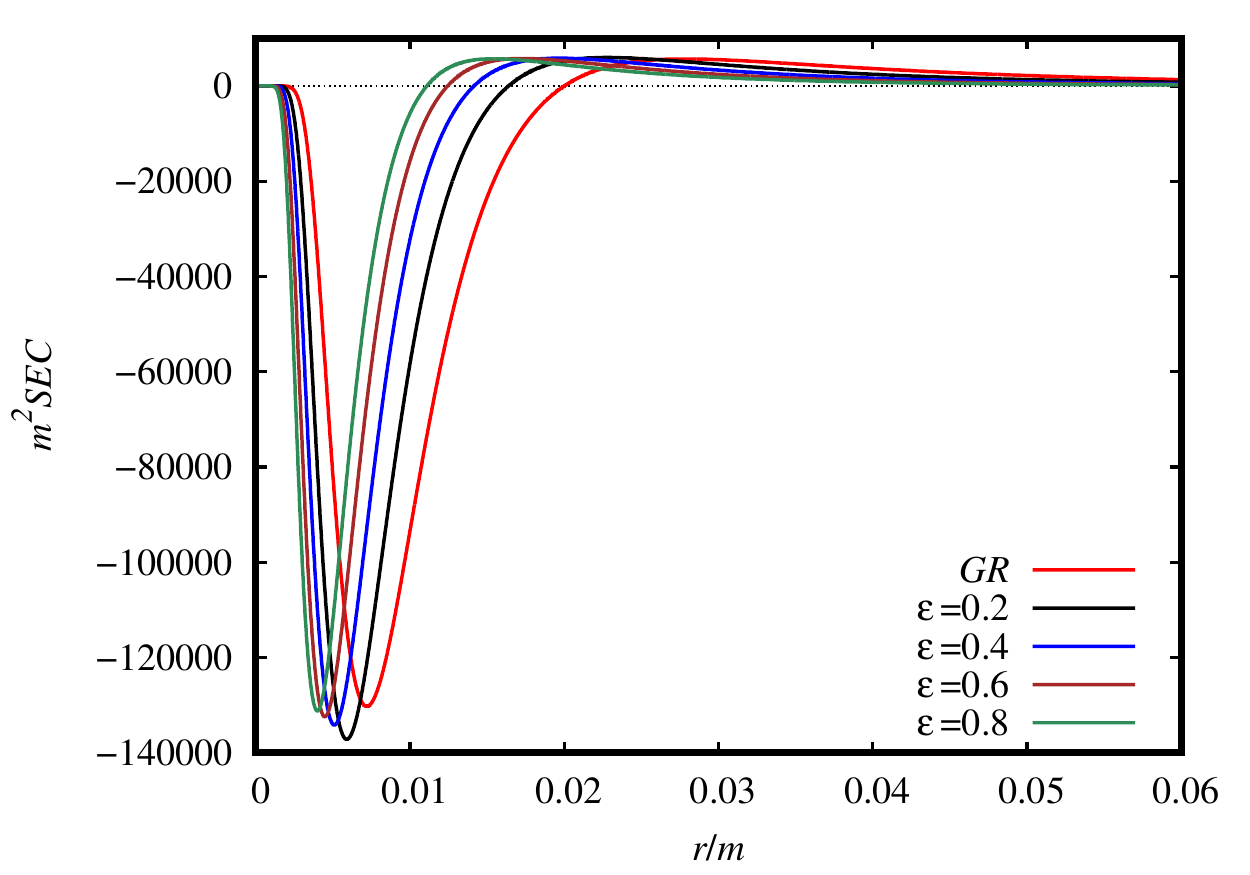}&\includegraphics[height=5cm,width=8cm]{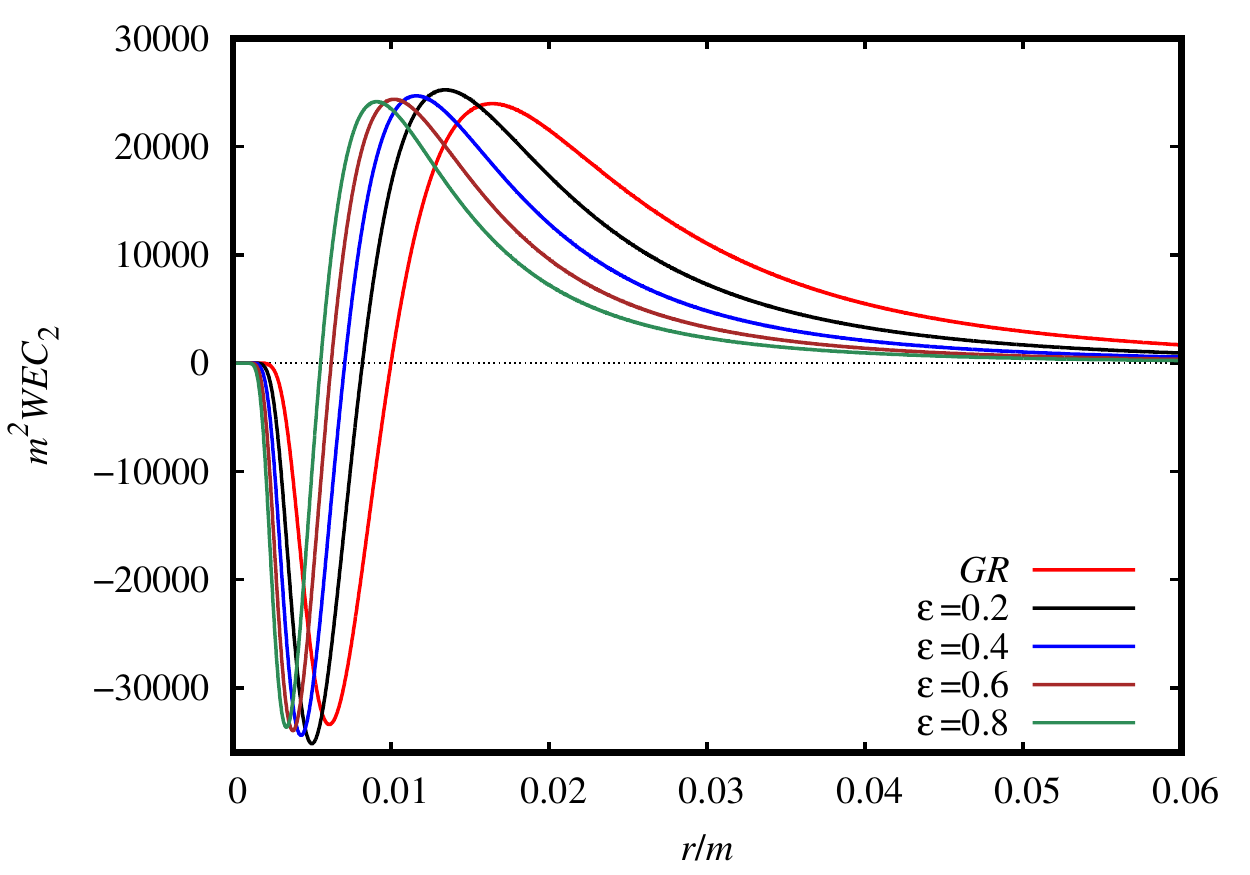}\\
\includegraphics[height=5cm,width=8cm]{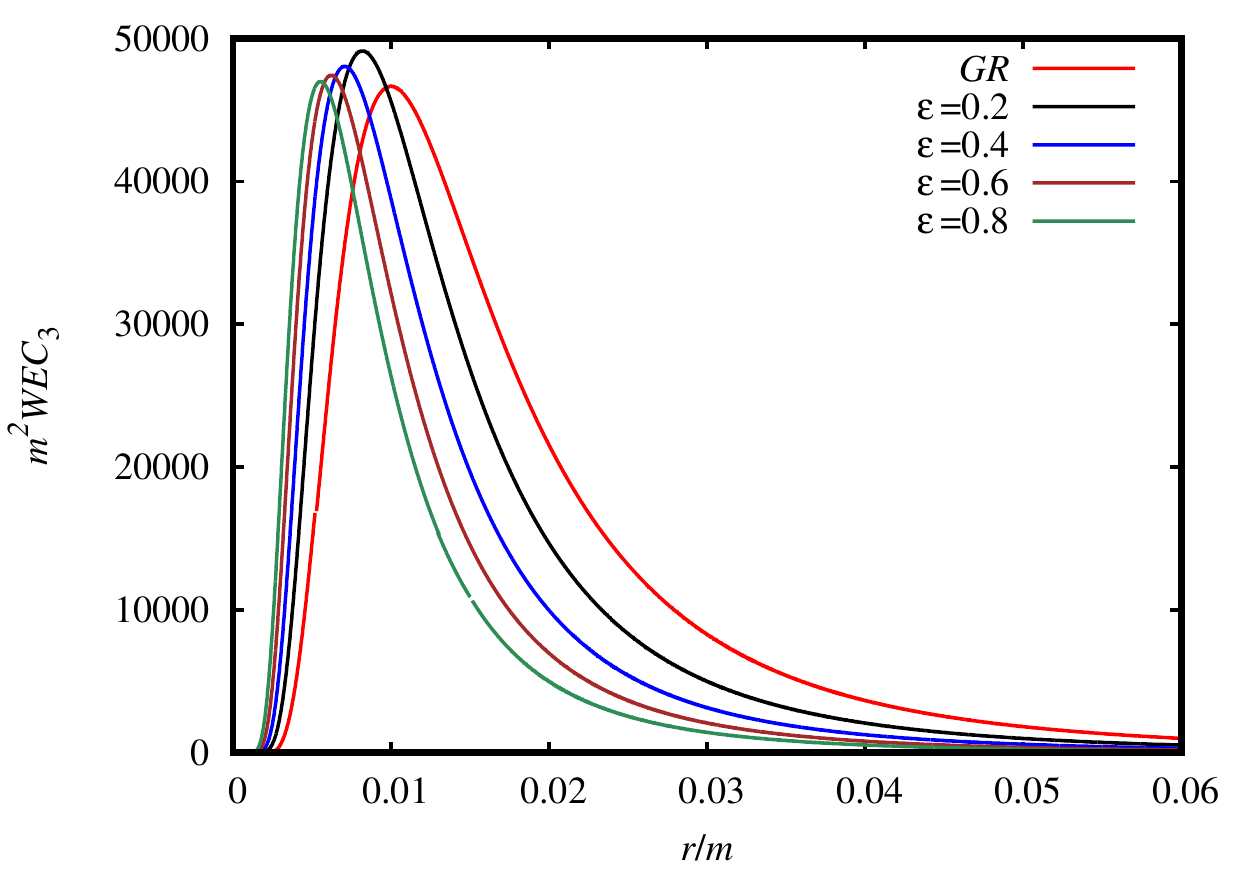}&\includegraphics[height=5cm,width=8cm]{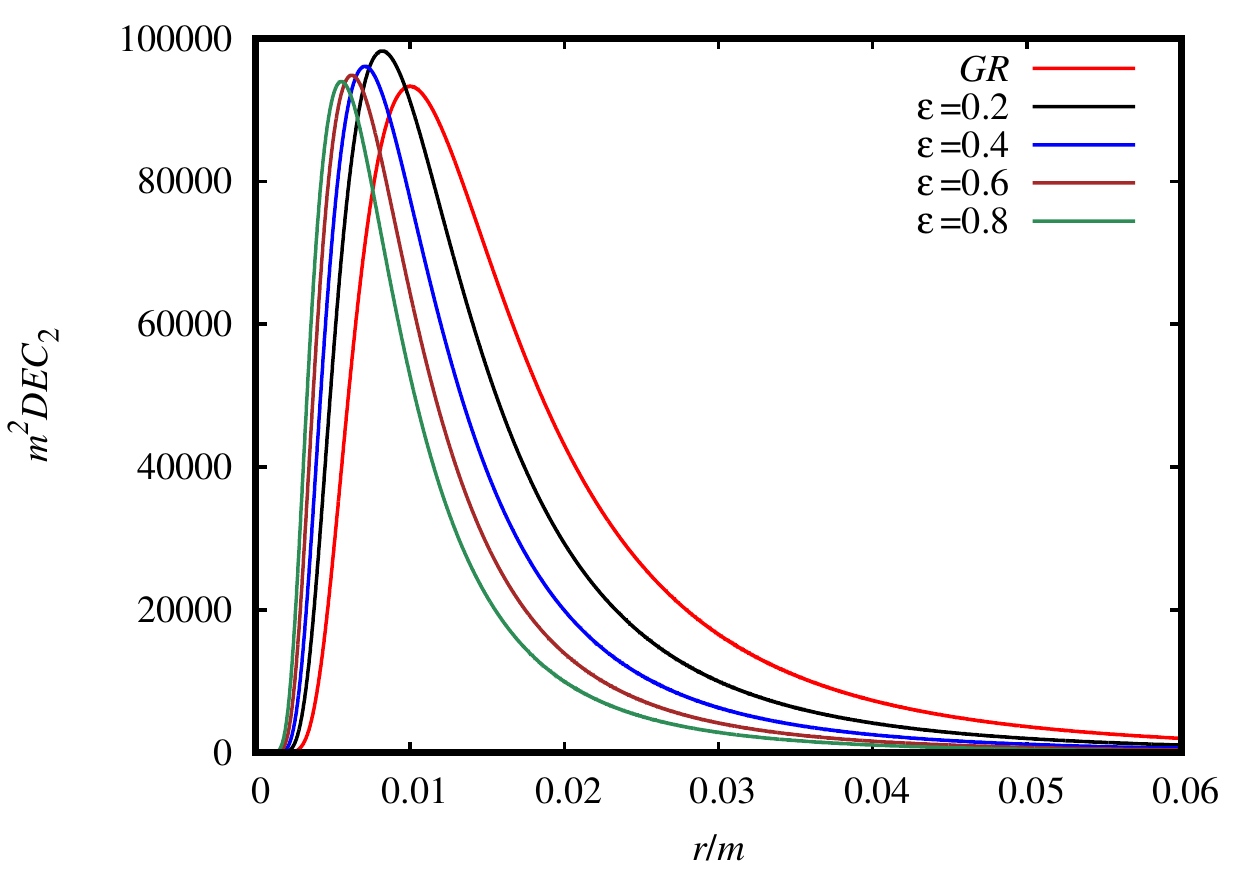}\\
\includegraphics[height=5cm,width=8cm]{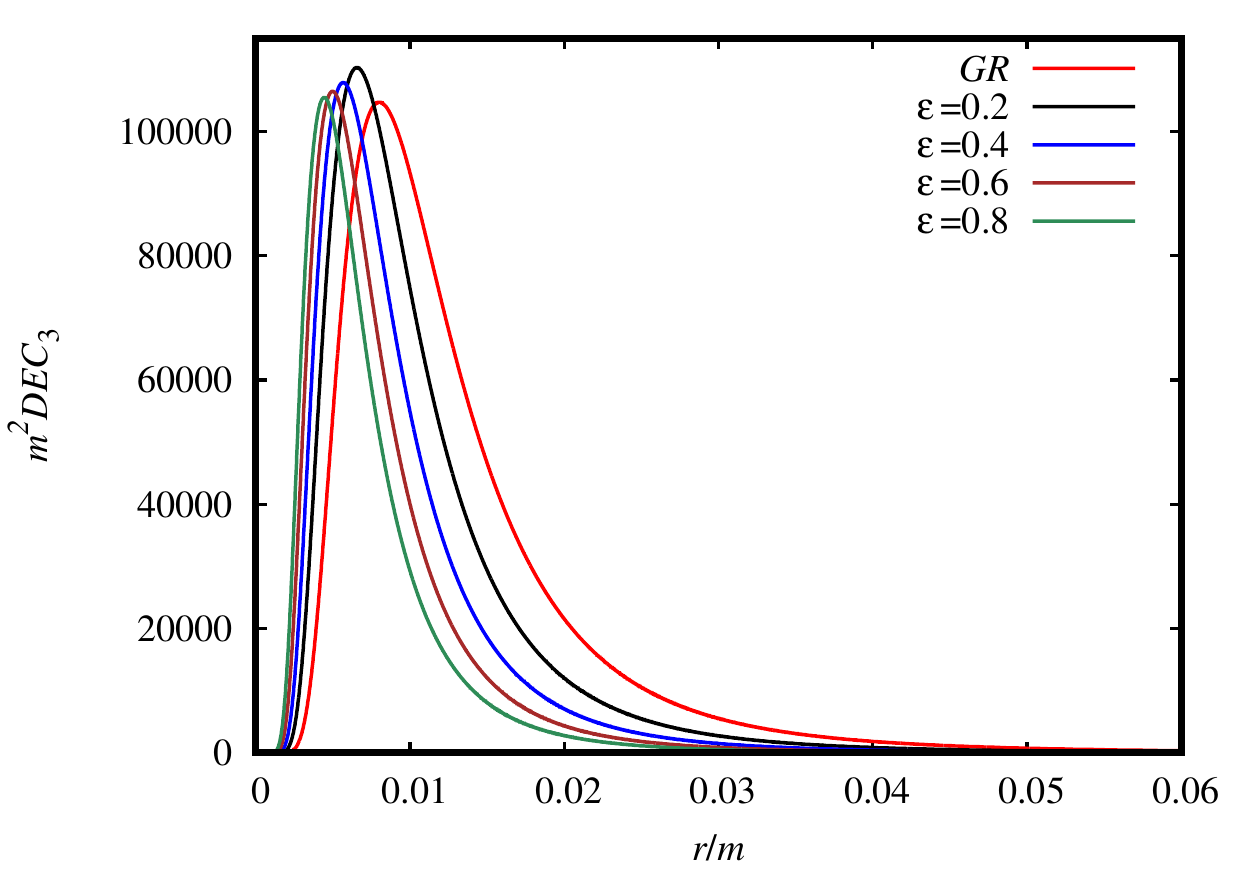}
\end{tabular}
\caption{\scriptsize{Graphical representation of the energy conditions SEC, WEC$_2$, WEC$_3$, DEC$_2$ e DEC$_3$ with $m=10Q$, $Q=0.8$, $\lambda=1$ and $G=1$ for some values of the energy $\epsilon$. We also see the GR case.} }
\label{fig4}
\end{figure}

Using the function $M(r,\epsilon)$ to the model that we are considering in \eqref{SEC}-\eqref{DEC} with \eqref{rho}, \eqref{pr1} and \eqref{pr2}, we get 
\begin{eqnarray}
&&SEC=\frac{e^{-\frac{Q^2g(\epsilon)k(\epsilon)}{2mr}}Q^2\,g^6(\epsilon)k(\epsilon)(4mr-Q^2\,g(\epsilon)k(\epsilon))}{16m\,\pi\,r^5 f^2(\epsilon)}\,,\\
&&WEC_1=0\,,\\
&&WEC_2=\frac{e^{-\frac{Q^2g(\epsilon)k(\epsilon)}{2mr}}Q^2\,g^6(\epsilon)k(\epsilon)(8mr-Q^2g(\epsilon)k(\epsilon))}{32m\,\pi\,r^5f^2(\epsilon)}\,,\\
&&WEC_3=\frac{e^{-\frac{Q^2g(\epsilon)k(\epsilon)}{2mr}}Q^2\,g^6(\epsilon)k(\epsilon)}{8\,\pi r^4 f^2(\epsilon)}\,,\\
&&DEC_2=\frac{e^{-\frac{Q^2g(\epsilon)k(\epsilon)}{2mr}}Q^2\,g^6(\epsilon)k(\epsilon)}{4\,\pi r^4 f^2(\epsilon)}\,,\\
&&DEC_3=\frac{e^{-\frac{Q^2g(\epsilon)k(\epsilon)}{2mr}}Q^4\,g^7(\epsilon)k(\epsilon)}{32\,m\pi r^5 f^2(\epsilon)}\,.
\end{eqnarray}
The energy conditions are represented in Fig. \ref{fig4}. All conditions are satisfied to $r\geq Q^2g(\epsilon)k(\epsilon)/4m$ inside the event horizon SEC and WEC are violated. The region where these functions are violated is attenuated as $\epsilon$ increases. The region where WEC is violated is smaller them to the SEC, since we need $r\geq Q^2g(\epsilon)k(\epsilon)/8m$ to satisfy the fist one.

\subsection{Balart-Vagenas-type solution}

\begin{figure}[h]
\centering
\begin{tabular}{rl}
\includegraphics[height=5cm,width=8cm]{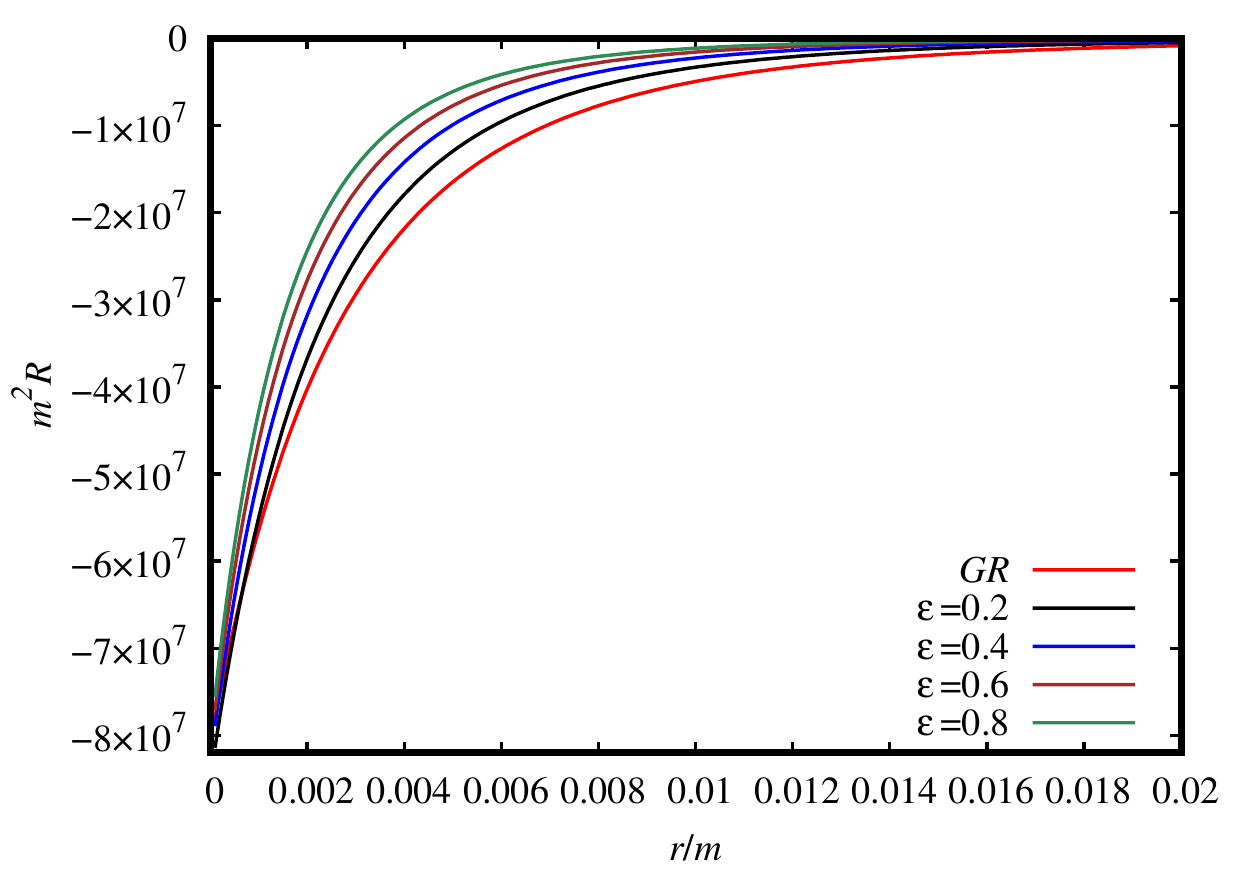}&\includegraphics[height=5cm,width=8cm]{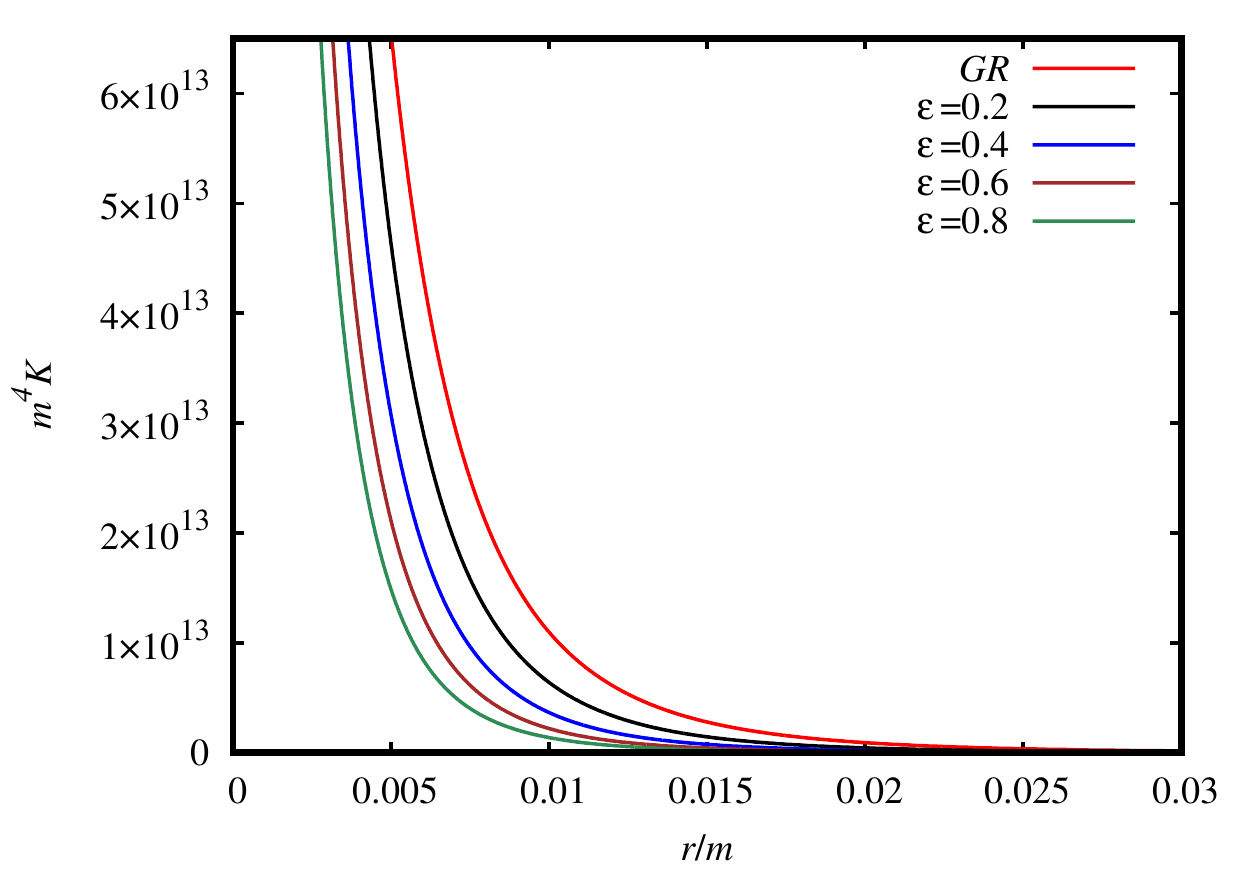}\\
\end{tabular}
\caption{\scriptsize{Scalars $R(r,\epsilon)$ and $K(r,\epsilon)$ to $m=10Q$, $Q=0.8$, $\lambda=1$ and $G=1$ for some values of $\epsilon$ and to GR.} }
\label{fig5}
\end{figure}

\begin{figure}[h]
\centering
\begin{tabular}{rl}
\includegraphics[height=5cm,width=8cm]{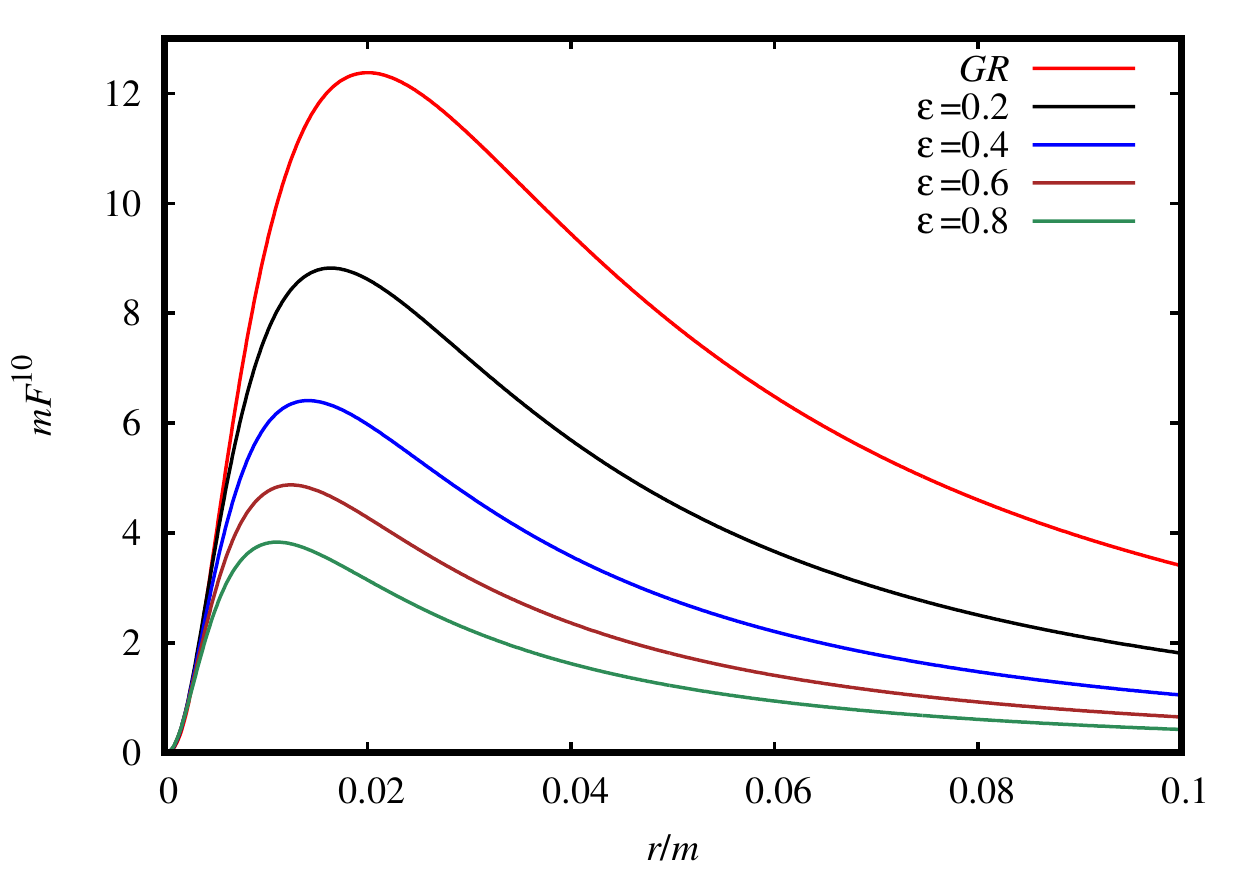}
\end{tabular}
\caption{\scriptsize{Electric field intensity \eqref{E102} with $m=10Q$, $Q=0.8$, $\lambda=1$, $G=1$ $\epsilon=0.2$, $\epsilon=0.4$, $\epsilon=0.6$ and $\epsilon=0.8$. The red line represent the GR case.} }
\label{fig52}
\end{figure}

Let's consider the mass function proposed in \cite{balart} in GR, where we make $M(r)\rightarrow M(r,\epsilon)$.
\begin{eqnarray}
M(r, \epsilon)=m\left(1+\frac{Q^2k(\epsilon)g(\epsilon)}{6mr}\right)^{-3}\label{vag}.
\end{eqnarray}
To $k(\epsilon)=g(\epsilon)=1$, we recover the GR case. As we did with the solution before, 
\begin{eqnarray}
&&e^{a(r)}=e^{-b(r)}=1-\frac{432\,G_N\,m^4 r^2\,f(\epsilon)^2h(\epsilon)^2}{g(\epsilon)(6mr+Q^2\,g(\epsilon)k(\epsilon))^3}\,,\label{eab2}\\
&&F^{10}(r,\epsilon)=\frac{243m^5\,Q\,r^3\,f(\epsilon)^3g(\epsilon)k(\epsilon)^{1/2}}{G_N\,\pi^2\,h(\epsilon)^2\left(6mr+Q^2\,g(\epsilon)k(\epsilon)\right)^5} \,,\label{E102}
\end{eqnarray} 
and the curvature invariants $R(r,\epsilon)$ and $K(r,\epsilon)$ are
\begin{eqnarray}
R(r,\epsilon)&=&-\frac{5184\,G_N\,m^4\,Q^4\,f(\epsilon)^2g(\epsilon)^3h(\epsilon)^2k(\epsilon)^2}{(6mr+Q^2g(\epsilon)k(\epsilon))^5}\,,\label{R2}\\
K(r,\epsilon)&=&\frac{4478976\,G^2_Nm^8\,f(\epsilon)^4g(\epsilon)^2h(\epsilon)^4}{(6mr+Q^2g(\epsilon)k(\epsilon))^{10}}\nonumber\\
&&\times\left[648m^4r^4+Q^2g(\epsilon)k(\epsilon)\left(126m^2Q^2r^2g(\epsilon)k(\epsilon)+Q^6g(\epsilon)^3k(\epsilon)^3-216m^3r^3\right)\right],
\end{eqnarray}
that are always regular. In the black hole center we find
\begin{equation}
\lim_{r\rightarrow 0}\left\{R(r,\epsilon),K(r,\epsilon)\right\}=\left\{-5184G_Nm^4f(\epsilon)^2h(\epsilon)^2/Q^6g(\epsilon)^2k(\epsilon)^3,4478976G_N^2m^8f(\epsilon)^4h(\epsilon)^4/Q^12g(\epsilon)^4k(\epsilon)^6\right\},
\end{equation}
which guarantees the regularity. We can also see the regularity in Fig. \ref{fig5} since we have no divergence in all spacetime. In the infinity of the radial coordinate the invariants tend to zero, so that, the solution is asymptotically flat. In Fig. \ref{fig52} we see ha behaves the intensity of the electric field and it's clear the regularity and that the intensity decreases as $\epsilon$ increases.

\begin{figure}[h]
\centering
\begin{tabular}{rl}
\includegraphics[height=5cm,width=8cm]{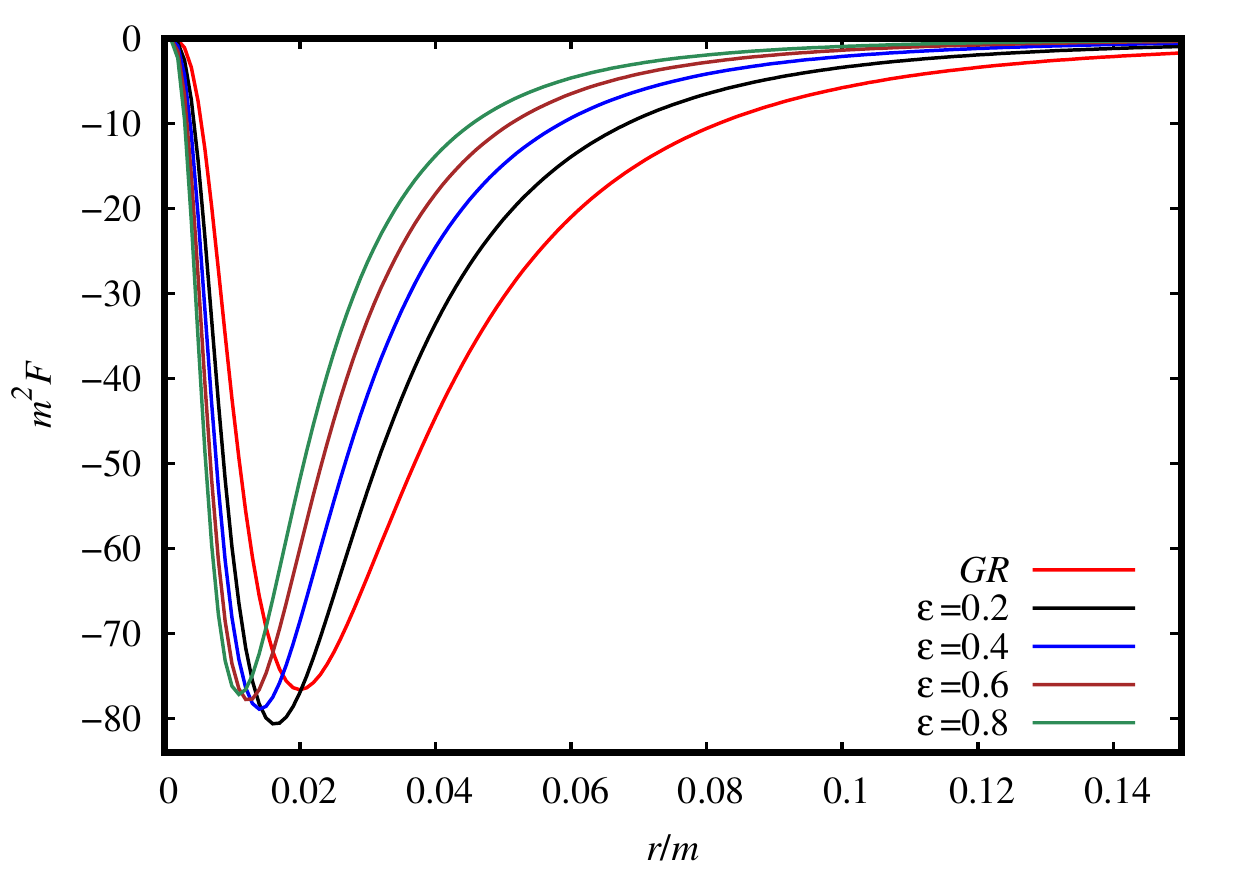}&\includegraphics[height=5cm,width=8cm]{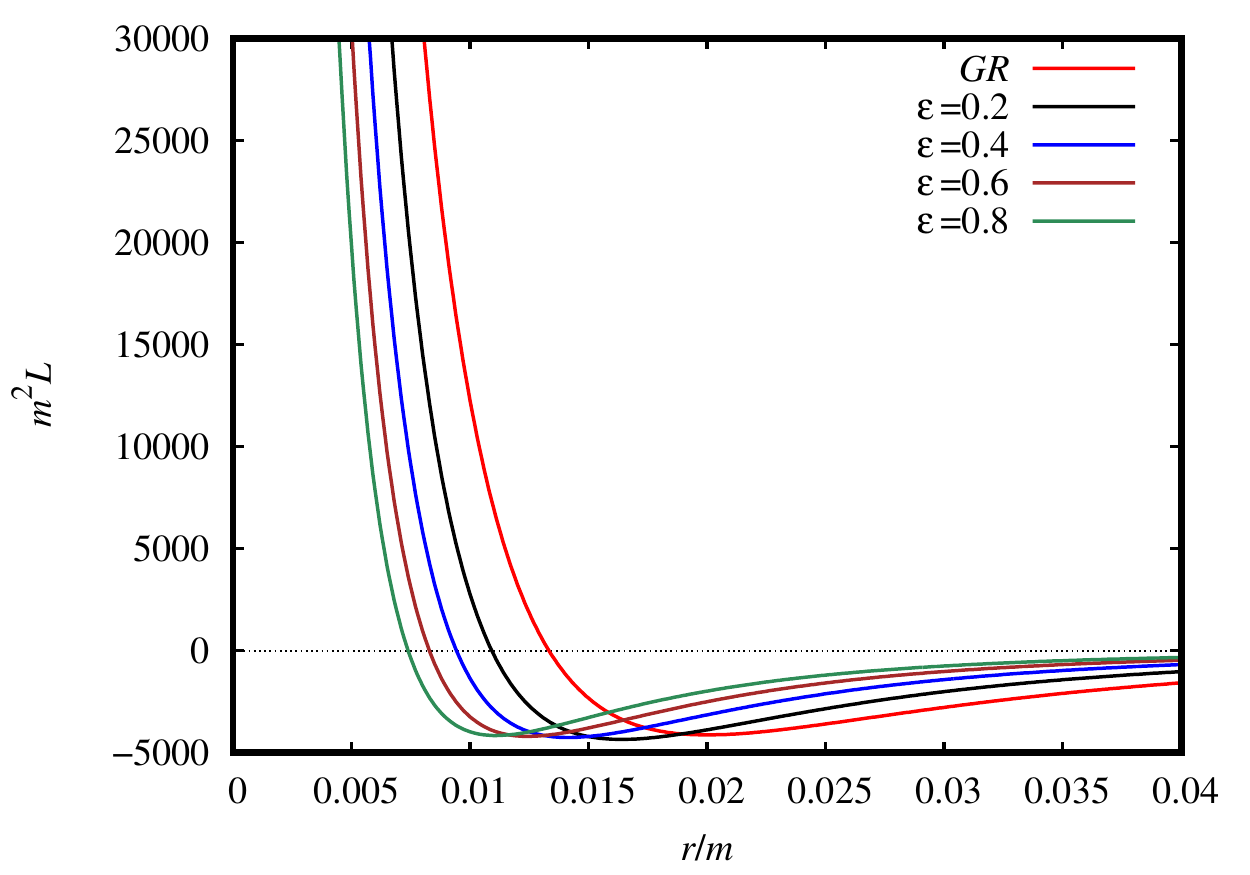}\\
\includegraphics[height=5cm,width=8cm]{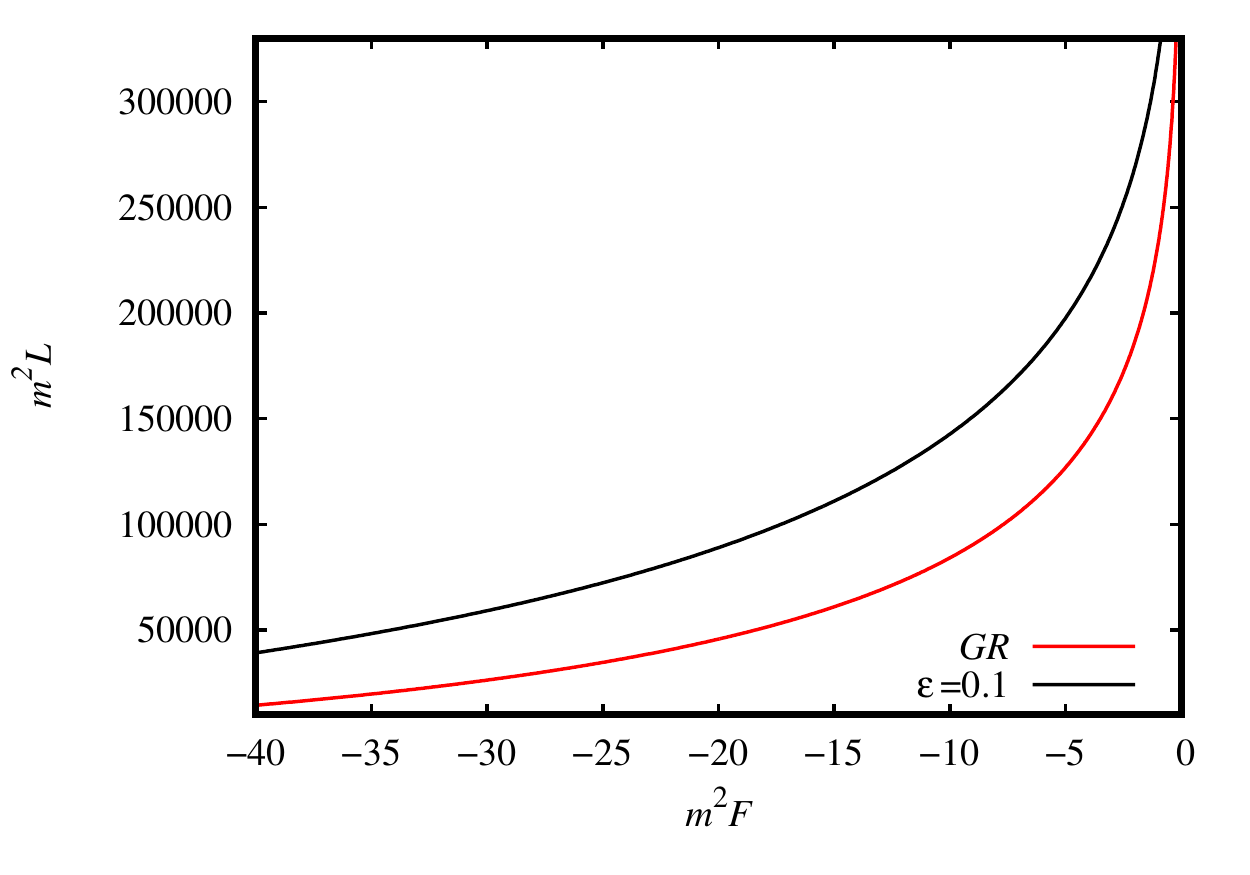}
\end{tabular}
\caption{\scriptsize{Graphical representation of $F(r,\epsilon)$(left) and $L(r,\epsilon)$(right) to $m=10Q$, $Q=0.8$, $\lambda=1$ and $G=1$ with $\epsilon=0.2$ $\epsilon=0.4$, $\epsilon=0.6$, $\epsilon=0.8$ and GR. The bottom graph show the behavior of $L(F)$ to $\epsilon=0.1$ and GR. } }
\label{fig7}
\end{figure}

We may also show the behavior of $F(r,\epsilon)$ and $L(r,\epsilon)$ in terms od the radial coordinate and compare to the GR case, as well the nonlinearity on the electromagnetic theory through $L(F)$, Fig \ref{fig7}.

The electromagnetic scalar and Lagrangian density are always regular, and the nonlinearity of $L(F)$ is evident, as it's expect to this type of solution.

\begin{figure}[h]
\centering
\begin{tabular}{rl}
\includegraphics[height=5cm,width=8cm]{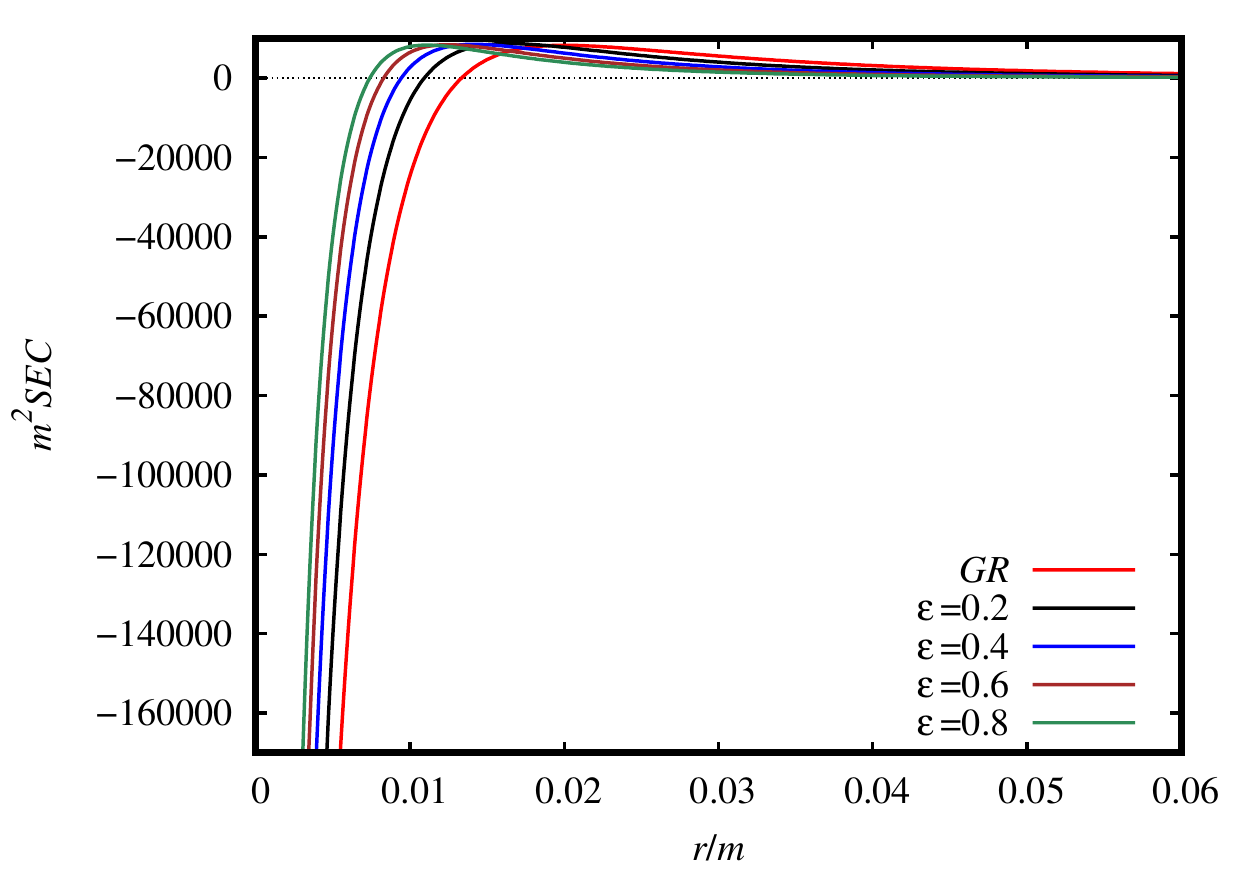}&\includegraphics[height=5cm,width=8cm]{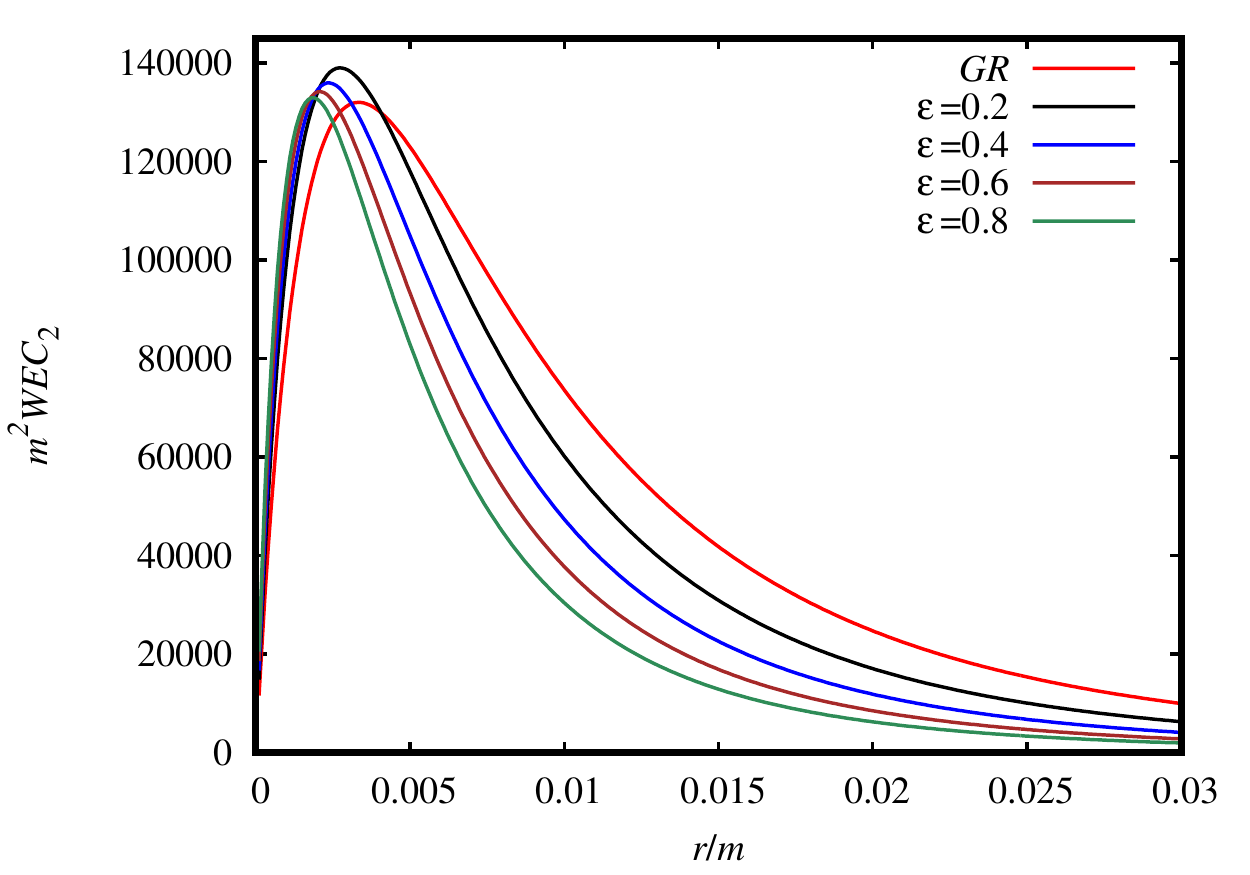}\\
\includegraphics[height=5cm,width=8cm]{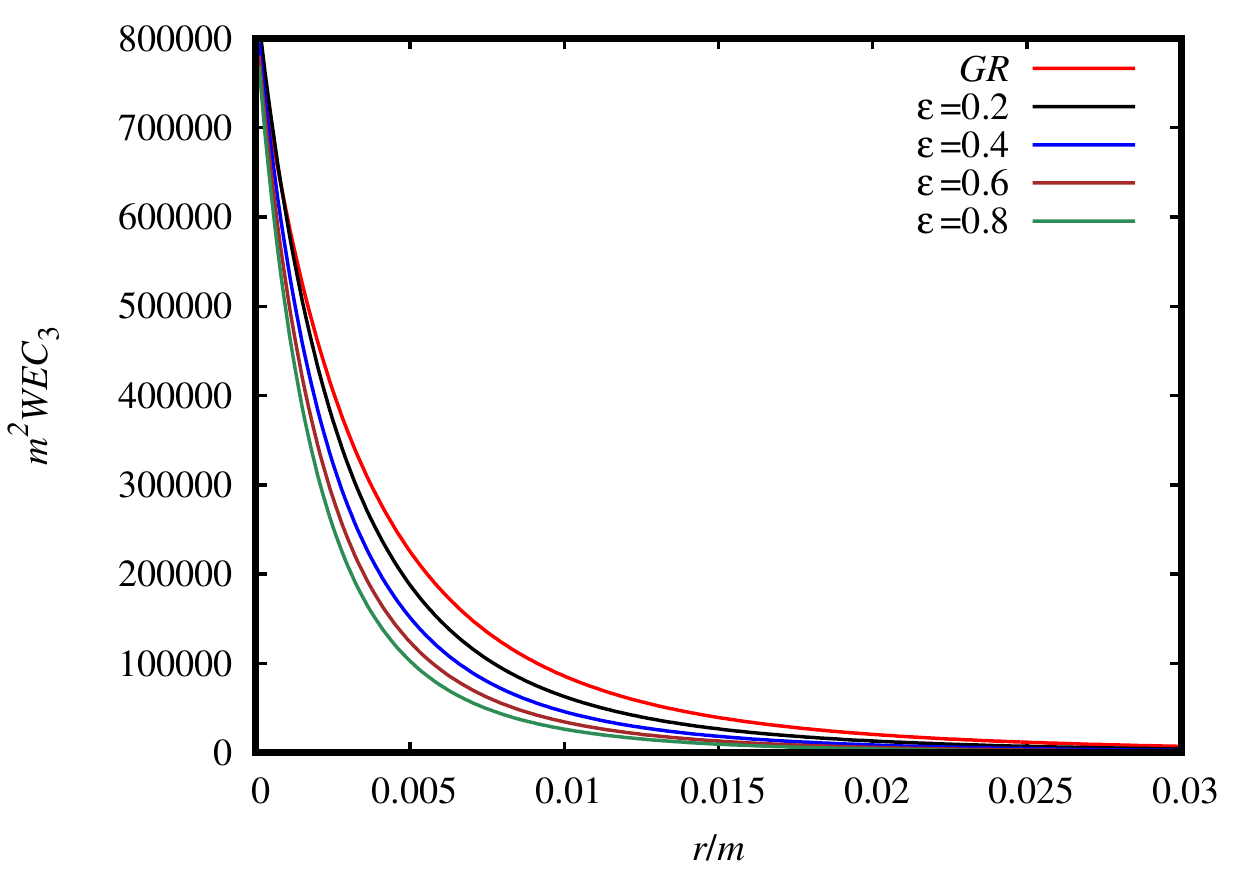}&\includegraphics[height=5cm,width=8cm]{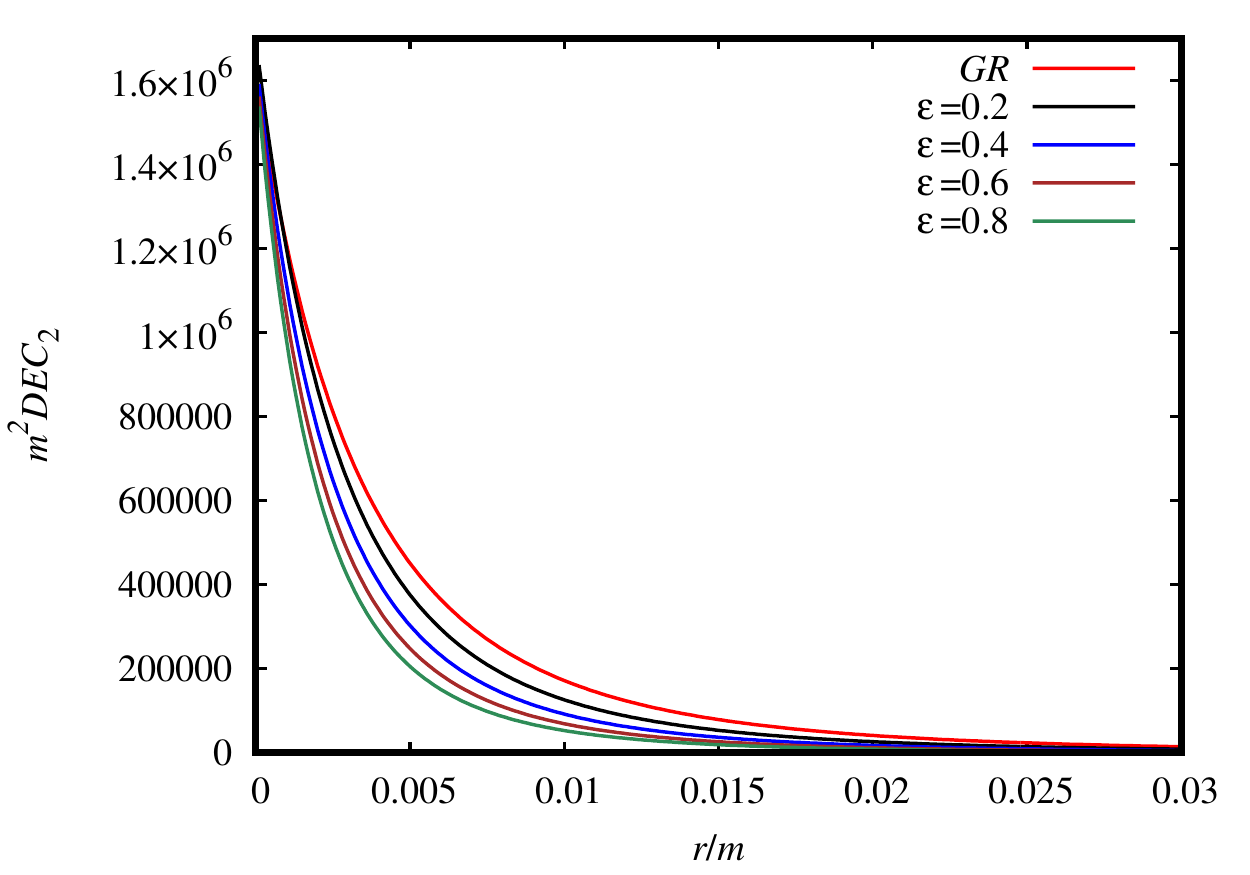}\\
\includegraphics[height=5cm,width=8cm]{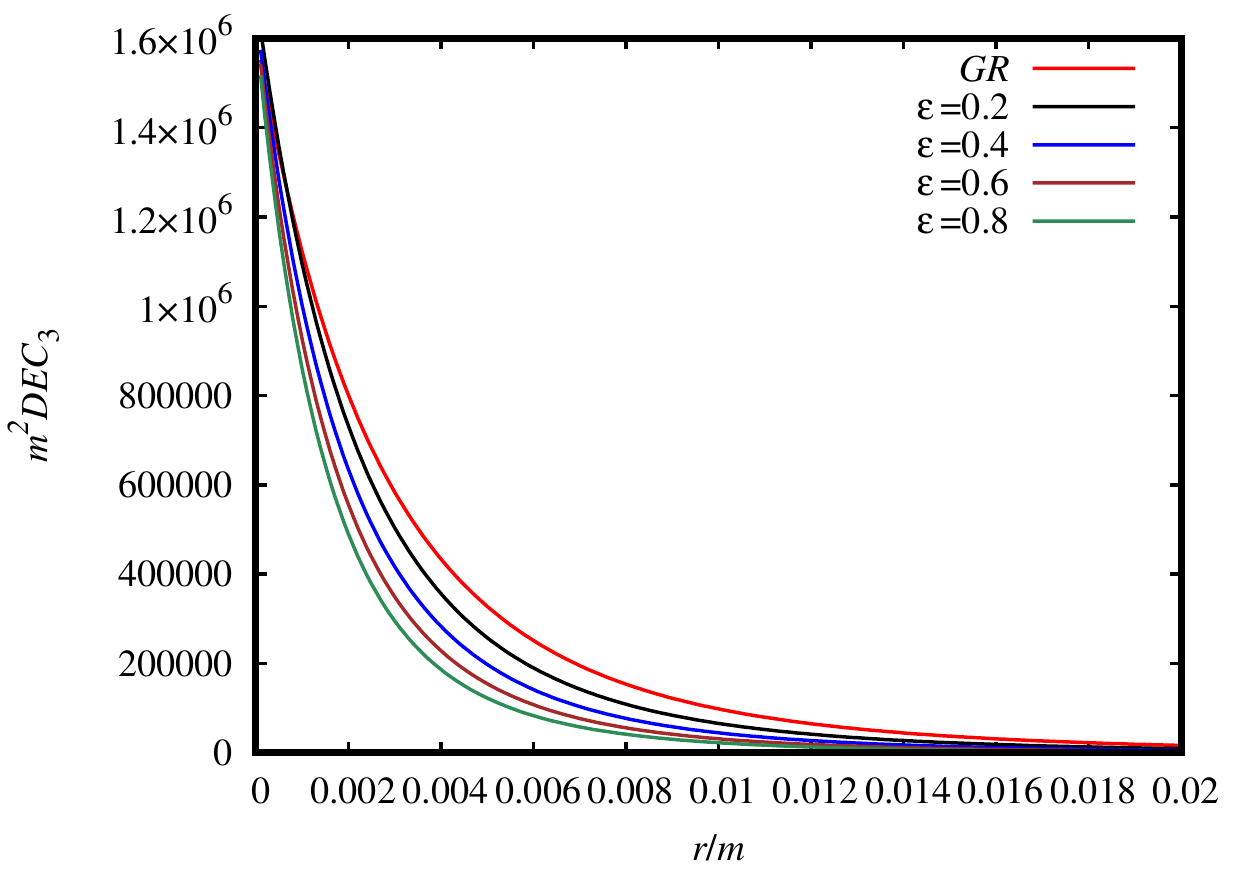}
\end{tabular}
\caption{\scriptsize{Energy conditions SEC, WEC$_2$, WEC$_3$, DEC$_2$ and DEC$_3$ with $m=10Q$, $Q=0.8$, $\lambda=1$ and $G=1$ for some of values of $\epsilon$. The red line represents the GR case.} }
\label{fig6}
\end{figure}

The energy conditions are found through \eqref{vag} and \eqref{SEC}-\eqref{DEC}, so that, we have
\begin{eqnarray}
&&SEC=\frac{324m^4Q^2g(\epsilon)^6k(\epsilon)(6mr-Q^2g(\epsilon)k(\epsilon))}{\pi f(\epsilon)^2(6mr+Q^2g(\epsilon)k(\epsilon))^5}\,,\label{Sec2}\\
&&WEC_1=0,\\
&&WEC_2=\frac{1944m^5Q^2rg(\epsilon)^6k(\epsilon)}{\pi f(\epsilon)^2(6mr+Q^2g(\epsilon)k(\epsilon))^5}\,,\\
&&WEC_3=\frac{162m^4Q^2g(\epsilon)^6k(\epsilon)}{\pi f(\epsilon)^2(6mr+Q^2g(\epsilon)k(\epsilon))^4}\,,\\\
&&DEC_2=\frac{324m^4Q^2g(\epsilon)^6k(\epsilon)}{\pi f(\epsilon)^2(6mr+Q^2g(\epsilon)k(\epsilon))^4}\,,\\
&&DEC_3=\frac{324m^4Q^4g(\epsilon)^7k(\epsilon)^2}{\pi f(\epsilon)^2(6mr+Q^2g(\epsilon)k(\epsilon))^5}\,.\label{D2}
\end{eqnarray}
The behavior of \eqref{Sec2}-\eqref{D2}, as a functions of $r$, is exhibit in Fig. \ref{fig6}. As in GR, to the model \eqref{vag} \cite{balart}, the conditions NEC, WEC and DEC are always satisfied. However, SEC is violated to $r<Q^2g(\epsilon)k(\epsilon)/6m$, inside the event horizon, but as we increase the $\epsilon$, the region where this condition is violated is attenuated. This result show explicit that the energy scale may modify the energy conditions.

\section{Conclusion}\label{conclusion}

We study the extent of the doubly special relativity made by J. Magueijo and L. Smolin in \cite{Smolin} where a modification of Einstein equations is made through the parameter $\epsilon=E/E_p$, where the Planck scale $E_p$ separated the classical regime to from the quantum one, with $0<\epsilon<1$, through the Rainbow functions $f(\epsilon)$ and $g(\epsilon)$. In this theory, when $\epsilon\rightarrow 0$, the velocity of a massless particle, as a photon, goes to $c$, and the classical regime is recovered. In Rainbow Gravity, the geometry depend on the energy of the particle that is used to test it. In this sense, particles with different energy see different geometries with the same inertial frame, by have the same equivalence principle \cite{Espdupla}.  

We present some model of regular black hole solutions in the Rainbow Gravity. To that, we consider the coupled with NED and extend the mass functions from Culetu and Balart-Vagenas to $M(r,\epsilon)$ through the Rainbow functions. In the Culetu-type solution, we showed that $R(r,\epsilon)$ and Kretschmann $K(r,\epsilon)$ are regular and asymptotically flat to $0<\epsilon<1$. The electromagnetic Lagrangian density is nonlinear in the scalar $F$ and is regular in all spacetime, as well as the electric field. About the energy conditions, SEC and WEC are violated near to black hole center and to high values of energy $\epsilon=E/E_p$, smaller is the region where this happens. DEC is always satisfied. The Balart-Vagenas-type solution, as the example before, is curvature regular and the electromagnetic theory is nonlinear. To this model, only the strong energy condition is violated and the region where it is violated decreases as the energy increases. To both models $\epsilon$ attenuate the intensity of the electric field.

A possible continuation of this work would be the study of black hole thermodynamics associated to regular solutions in Rainbow Gravity, similar to what is found in the literature \cite{Panah}.


\vspace{1cm}

{\bf Acknowledgement}: M. E. R. thanks CNPq for partial financial support. The authors thanks André Carlos Lehum and Vitor Cardoso for helpful discussions. This study was financed in part by the Coordenação de Aperfeiçoamento de Pessoal de Nível Superior - Brasil (CAPES) - Finance Code 001. 

\end{document}